%% file: main-jss2.tex
\documentclass[a4paper,fleqn]{cas-dc}

\usepackage[authoryear,longnamesfirst]{natbib}
\usepackage{import}
\usepackage{makeidx}
\makeindex
\usepackage[utf8]{inputenc}
\usepackage[dvipsnames]{xcolor}
\usepackage{tikz, pgfplots}
\usepgfplotslibrary{statistics}
\usetikzlibrary{positioning}
\pgfplotsset{compat=1.18}
\usepackage{filecontents}

\newcommand*\circled[1]{\tikz[baseline=(char.base)]{
            \node[shape=circle,draw,inner sep=2pt] (char) {#1};}}

\newcommand*{\myindent}{\hspace*{0.5cm}}
\newcommand{\jssrev}[1]{#1}
\newcommand{\jssrevtwo}[1]{#1}

\definecolor{lightblue}{RGB}{32, 209, 228}
\definecolor{darkgreen}{RGB}{0, 64, 0}
\definecolor{darkyellow}{RGB}{255, 204, 0}

\def\tsc#1{\csdef{#1}{\textsc{\lowercase{#1}}\xspace}}
\tsc{WGM}
\tsc{QE}

\begin{document}
\let\WriteBookmarks\relax
\def\floatpagepagefraction{1}
\def\textpagefraction{.001}

\shorttitle{Studying the association between Gitcoin's issues and resolving outcomes}    

\shortauthors{M. Choetkietikul, A. Puengmongkolchaikit, P. Chandra, C. Ragkitwetsakul, R. Maipradit, H. Hata, T. Sunetnanta, K. Matsumoto}  

\title [mode = title]{Studying the association between Gitcoin's issues and resolving outcomes}

\author{Morakot Choetkiertikul}
\fnmark[a]
\ead{morakot.cho@mahidol.ac.th}

\author{Arada Puengmongkolchaikit}
\fnmark[a]
\ead{arada.pue@student.mahidol.ac.th}

\author{Pandaree Chandra}
\fnmark[a]
\ead{pandaree.cha@student.mahidol.ac.th}

\author{Chaiyong Ragkitwetsakul*}
\fnmark[a]
\ead{chaiyong.rag@mahidol.ac.th}

\author{Rungroj Maipradit}
\fnmark[b]
\ead{maipradit.rungroj.mm6@is.naist.jp}

\author{Hideaki Hata}
\fnmark[c]
\ead{hata@shinshu-u.ac.jp}

\author{Thanwadee Sunetnanta}
\fnmark[a]
\ead{thanwadee.sun@mahidol.ac.th}

\author{Kenichi Matsumoto}
\fnmark[b]
\ead{matumoto@is.naist.jp}

\affiliation[a]{organization={Faculty of Information and Communication Technology, Mahidol University}, Thailand}

\affiliation[b]{organization={Nara Institute of Science and Technology}, Japan}

\affiliation[c]{organization={Shinshu University}, Japan}

\cortext[1]{Corresponding author}



\begin{abstract}
The development of open-source software (OSS) projects usually \jssrev{have} been driven through collaborations among contributors and strongly relies on volunteering. Thus, allocating software practitioners (e.g., contributors) to a particular task is non-trivial and draws attention away from the development. Therefore, a number of bug bounty platforms have emerged to address this problem through bounty rewards. Especially, Gitcoin, a new bounty platform, introduces a bounty reward mechanism that allows individual issue owners (backers) to define a reward value using cryptocurrencies rather than using crowdfunding mechanisms. Although a number of studies have investigated the phenomenon on bounty platforms, those rely on different bounty reward systems. Our study thus investigates the association between the Gitcoin bounties and their outcomes (i.e., success and non-success). We empirically study over 4,000 issues with Gitcoin bounties using statistical analysis and machine learning techniques. \jssrev{We also conducted a comparative study with the Bountysource platform to gain insights into the usage of both platforms. Our study highlights the importance of factors such as the length of the project, issue description, type of bounty issue, and the bounty value, which are found to be highly correlated with the outcome of bounty issues. These findings can provide useful guidance to practitioners.}
\end{abstract}




\maketitle

\section{Introduction}
\label{section-intro}

    The development of Open-Source Software (OSS) projects  is encouraged through collaborations and knowledge sharing among project contributors. 
    It mostly relies on volunteer work. Recruiting an experienced developer for a challenging task becomes a non-trivial task \citep{Choi2010}. 
    Thus, bounty platforms are used to support the development of OSS projects through the mechanisms of bounty and crowdsourcing \citep{Zhou2021, Kanda2017, Zhou2020, Hata2017}. 
    A bounty refers to a reward for developers who can accomplish tasks, such as bug fixing or new feature creation, declared on a bounty platform \citep{Finifter2013}. 
    Those rewards could be in cash, cryptocurrency, or badges that can reflect the developer's reputation \citep{Nakasai2019}. 
    The previous study has shown that monetary incentive is an essential motivation that can attract the developers to contribute to each project on the platforms \citep{Hata2017}.

    Recently, there have been a number of bounty platforms allowing contributors to participate and contribute to OSS projects, such as Bountysource\footnote{\url{https://www.bountysource.com}} and 
    HackerOne.\footnote{\url{https://www.hackerone.com}} Specifically, the crowdsourcing bounty platform allows backers to support any specific tasks of projects, such as implementing new features or fixing bugs by providing 
    rewards on resolving tasks as bounties. Several empirical studies aim to investigate the impact on software development projects using this bounty mechanism. 
    \cite{Kanda2017} report that bounties tend to attract developers to work on the projects more than those without bounties. 
    Several existing empirical studies on the Bountysource platform found that the amount of bounties is a factor that affects the outcome of the project \citep{Kanda2017, Zhou2020}. 
    
    Gitcoin\footnote{\url{http://gitcoin.co}} is a new bounty platform established in 2017. 
    It plays a role as a bounty-based collaboration community for funders (i.e., bounty issue owners) and developers (i.e., contributors) to work on the issues easily on GitHub.\footnote{\url{https://gitcoin.co/bounties/funder}} 
    Gitcoin purely uses the Ethereum blockchain and cryptocurrency, such as Bitcoin, for their reward system. 
    Funders provide a bounty in cryptocurrency-based rewards. 
    In particular, Gitcoin allows a funder to solely support a bounty task, 
    such as a project development task and bug resolving task, which is different from traditional bounty platforms that rely on crowdfunding mechanisms 
    that require a number of funders to support a bounty. Currently, Gitcoin is gaining more attention from funders and 
    developers because of its unique reward system, as the number of bounties created on Gitcoin and the number of active developers are increasing. 
    Over 6,000 bounty issues have been created on Gitcoin, and over 10,000 active developers participate in those bounties.\footnote{The data were collected on 20 January 2021.} 
    \jssrevtwo{While previous studies have focused on traditional bounty platforms such as Bountysource, it is important to note that these platforms share some common characteristics while also exhibiting differences. 
    there remains a gap in understanding the dynamics and characteristics specific to cryptocurrency-based bounty platforms. 
    To address this gap, a study on cryptocurrency-based bounty platforms and a comparative study between the two platforms are required.}
    
    
    


    In this paper, we  aim to investigate the factors that impact the success of the bounty issue resolution created in Gitcoin. We thus adopt the studying approach from \cite{Zhou2020,Zhou2021}. 
    In our study, we empirically study over 4,000 bounty issues (i.e., issue reports) in Gitcoin created between September 2017 to December 2020, which involve 1,096 software project repositories hosted in GitHub with a total bounty value of over 18,000 ETH and 11,000,000 USDT. Our study extracted \jssrevtwo{28} features characterizing the Gitcoin bounty issues (e.g., developer's experience level, bounty proposing time, and estimated development time). \jssrevtwo{The study conducted by \cite{Zhou2020,Zhou2021} on Bountysource highlighted the significant relationship between the likelihood of issue resolution and factors such as the timing of proposing bounties, the bounty value, and the duration of the issue. Our study, conducted on a different bounty platform, confirms these findings as we observe similar factors that impact issues on Gitcoin. However, our study also uncovers additional insights. We found that the description of tasks, the experience level of contributors, and the type of contribution are crucial in determining the success rate of issues. Furthermore, our comparative analysis reveals differences in terms of project types and focused topics on these platforms, which can serve as a guide for platform selection and creating issues with a higher chance of success. } 

    

    This paper is organized as follows. Section \ref{section-background} provides the background of bounty platforms, explains the Gitcoin platform, and discusses related work on bounty platform studying. Section \ref{section-method} discusses our studying approach, including data collection, preprocessing and labeling, and data analysis. We then discuss our extracted features of the Gitcoin bounties in Section \ref{section-feature}. In Section \ref{section-correlation}, we discuss the results from our feature and correlation analysis. We then report and discuss our findings in identifying important features in Section \ref{section-classification}.  In Section \ref{section-tov}, we discuss the threats to the validity of our study, and we conclude our study in Section~\ref{section-conclusion}.

\section{Background and related work}
\label{section-background}

\subsection{Bounty platforms}
\label{subsection-bountyplatform}
    Using bounties to accomplish tasks in software development projects has become a popular approach \citep{Atiq2014}. Since an OSS project allows developers to participate in the development of the projects, a bounty can thus enlarge the engagement of developers to work on the projects \citep{Zhou2020a}. A bounty could be in various forms, such as cash, cryptocurrency, and a badge. A badge is used to enhance the reputation of developers \citep{Nakasai2019}. Funding the project as a bounty is a popular way to drive OSS projects, which allows any developers to work on the projects freely.
    A funder can be either an individual or an organization that needs to accomplish a specific task \citep{Coelho2018}. For example, they are a project user and need to have a new feature implemented in the project.

    Several studies were conducted to understand both intrinsic and extrinsic incentives of developer contributions. An intrinsic incentive refers to the enjoyment of contribution, the ability to work on the projects, or even the satisfaction of solutions \citep{Krishnamurthy2014}. In contrast, an extrinsic incentive can be motivated by the desire to get rewards. Studies reported that the developers are likely to complete tasks faster if an amount of bounty is provided \citep{Kanda2017, Hata2017}. The Bountysource platform is among the most popular and widely used \citep{Zhou2020}. Bountysource allows several funders to fund a bounty issue. Then, the developers can work on the projects through an issue tracking system such as GitHub and Bugzilla. In contrast, a Gitcoin bounty platform allows only one funder to fund a bounty issue.

    \begin{figure*}[h]
            \centering
            \includegraphics[width=12cm]{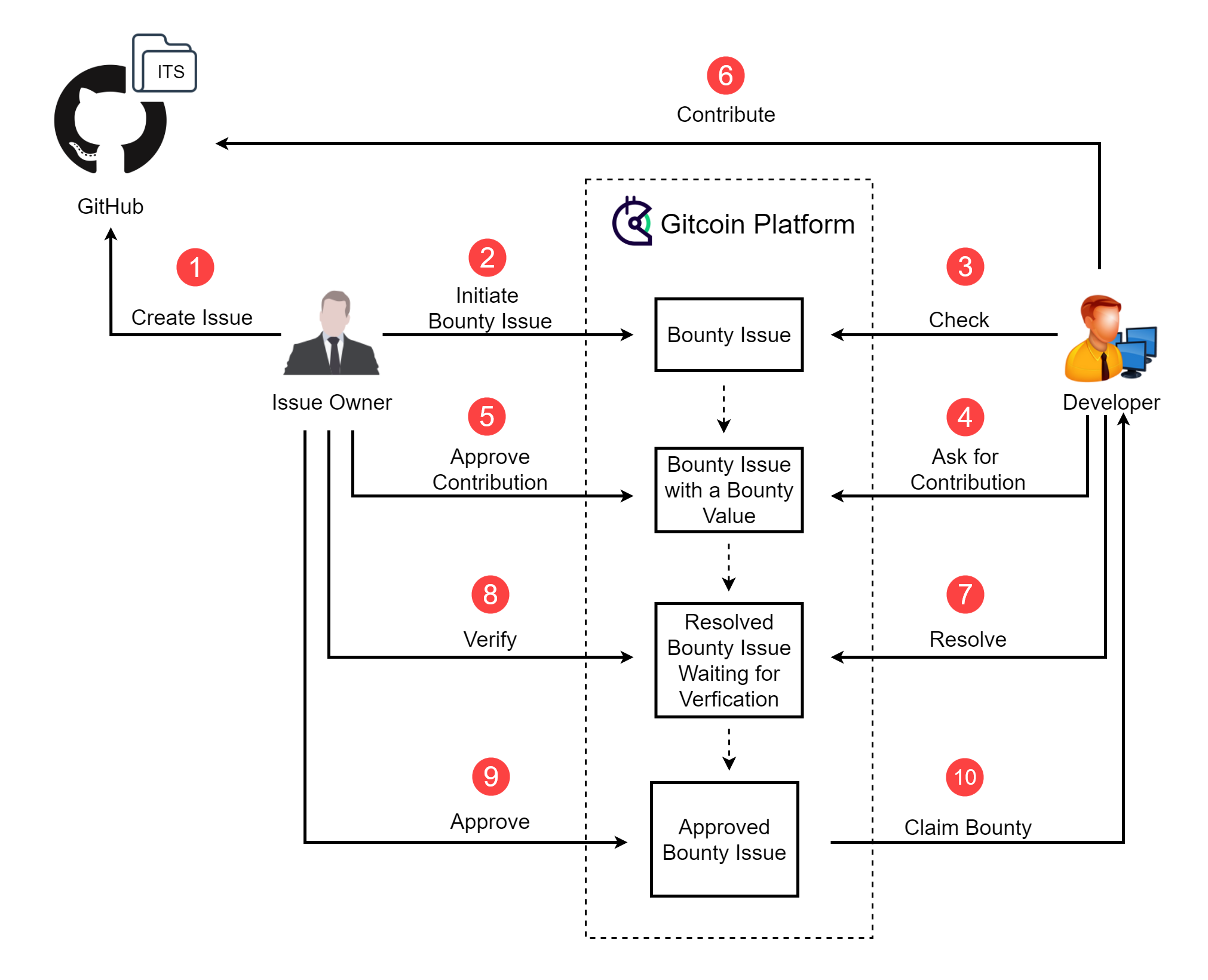}
            \caption{Gitcoin issue's lifecycle}
            \label{figure-gitcoinflow}
    \end{figure*}

\subsection{Gitcoin} 
\label{subsection-gitcoin}
    Gitcoin is a bounty platform that allows developers to contribute to OSS projects  with bounties. Gitcoin purely uses cryptocurrencies as a bounty reward for those who can resolve issues. This rewarding approach provides a high level of security on decentralized systems and no cost of transaction \citep{Titov2021}. Gitcoin supports several types of tokens in cryptocurrency networks, such as Ethereum (ETH) and Bitcoin (BTC) \citep{Li2019}. Gitcoin particularly supports ERC20 tokens in its ecosystem. ERC20 uses the Smart Contracts program to help organize and formalize the agreements on the networks to work properly among users (e.g., companies and entrepreneurs) \citep{Chen2020}. Therefore, advocating ERC20 on the Gitcoin platform could ensure the authenticity and credibility between funders and contributors. With this intention, the funders can use well-known cryptocurrencies to fund issues on Gitcoin safely in any granted types of tokens on the platform.

\begin{figure*}[htbp]
        \centering
        \includegraphics[width=12cm]{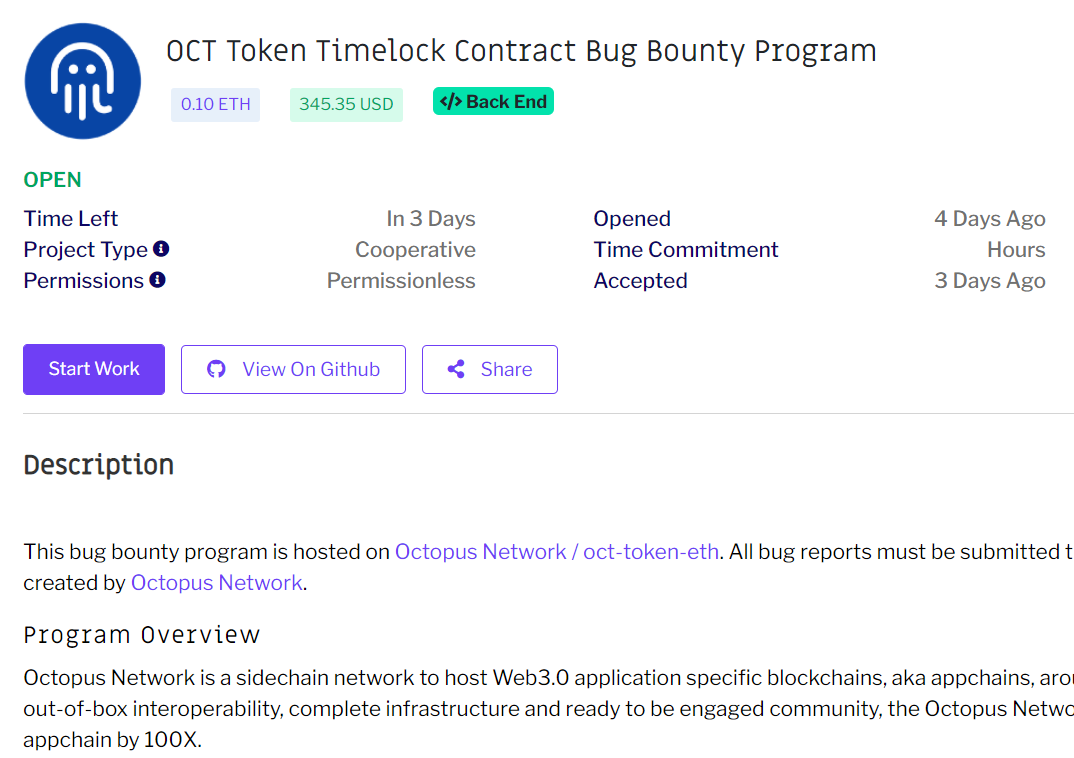}
        \caption[example]{An example of the Gitcoin issue\protect\footnotemark}
        \label{figure-gitcoinissue}
    \end{figure*}

    Figure \ref{figure-gitcoinflow} shows the lifecycle of the Gitcoin bounty issue. Gitcoin uses GitHub's issue-tracking system to track issue-resolving progress. A GitHub issue corresponding to the Gitcoin bounty issue must exist in the GitHub issue tracking system. The GitHub issue is usually initiated by an issue owner (i.e., funder/backer) of a GitHub repository in Step \circled{1}. The issue owner then initiates the issue on the Gitcoin platform and fills in the issue's details, such as title, \jssrev{contribution} type, required experience level, expected task duration, and bounty value, as well as the corresponding GitHub issue link (Step~\circled{2}).  In Step~\circled{3}, developers (i.e., contributors) can find Gitcoin issues that they would like to contribute from the Gitcoin platform. In addition, the issue owner can select whether developers must be approved before contributing or whether any developers can contribute to the issue (i.e., permissionless mode). In the former case, the prospect developer must send a request to contribute to the issue owner and wait for the acceptance before starting the work (\circled{4}). Once the developer has been approved to work on the issue (\circled{5}), they can access to source code and related resources from the provided GitHub issue and can also communicate with the issue owner from there. The developer resolves the issue in Step \circled{7}, and the issue owner can review the issue and decide whether to accept the issue resolution in Step \circled{8}. The issue owner then confirms the completion of the task on the Gitcoin platform to close the Gitcoin issue (\circled{9}). The developer can then claim the bounty reward in the last step (\circled{10}). Figure \ref{figure-gitcoinissue} shows an example of a Gitcoin issue.

\subsection{Studies of bounty platforms}
    Several bounty programs were launched in the past. For example, proprietary software like Google or Mozilla Firefox allowed internal employees to work on the vulnerability rewards programs \citep{Luna2019, Finifter2013}. The empirical study from \cite{Finifter2013} described two vulnerability rewards programs (VRPs) between Chrome and Mozilla Firefox. They indicated that the monetary incentive could also help prevent the researchers from selling the research results of the system's security in a gray market. In addition, \cite{Zhao2014} studied platforms that act as the middlemen between the contributors and the software vendors, such as BugCrowd,\footnote{\url{https://www.bugcrowd.com/}} Synack,\footnote{\url{https://www.synack.com/}} and CrowdCurity (as know as Cobalt at present).\footnote{\url{https://cobalt.io/}} In particular, bounties in these platforms focus on security and penetration testing. 
    

    

    \footnotetext{\url{https://gitcoin.co/issue/octopus-network/oct-token-eth/1/100026469}}

    Apart from the vulnerability rewards programs, the empirical studies of bounty platforms for open-source software projects are also ubiquitous in the software engineering research community. A number of projects cannot be accomplished without collaboration among the contributors in a community. According to the study from \cite{Zhou2020a}, a question-and-answer forum such as Stack Overflow\footnote{\url{https://stackoverflow.com/}} is one type of bounty mechanism that allows contributors (i.e., answerers) to get the forum's reputation points. Moreover, \cite{Wang2018} described that the incentive systems like gamification were developed to encourage users to provide answers to questions. Thus, it can help motivate users to engage with the questions. 
    

    As the number of bounty hunters (i.e., the ones who mainly work on resolving issues with a bounty reward) is increasing in software development communities, a crowdsourcing bounty platform allows multiple funders to fund a bounty. The empirical study of Bountysource by \cite{Kanda2017} demonstrated that issues with bounties are more likely to be solved than those projects without bounties.  Further empirical studies from \cite{Zhou2020} and \cite{Zhou2021} indicate that, on the Bountysource bounty platform, funders are likely to offer the bounties in higher value and more frequently than the individual backers \citep{Zhou2021}. They also reported that the bounty value is not the most important factor that attracts contributors to address the issues. They indicated that some contributors are not motivated by only rewards or monetization. Instead, they could be driven by their own interests or desires to commit to the work \citep{Zhou2020}. We thus adopt their studying approach to our study to understand the factors deriving from the Gitcoin bounty platform.

    

\section{Methodology}
\label{section-method}
\begin{figure*}[h]
        \centering
        \includegraphics[width=12cm]{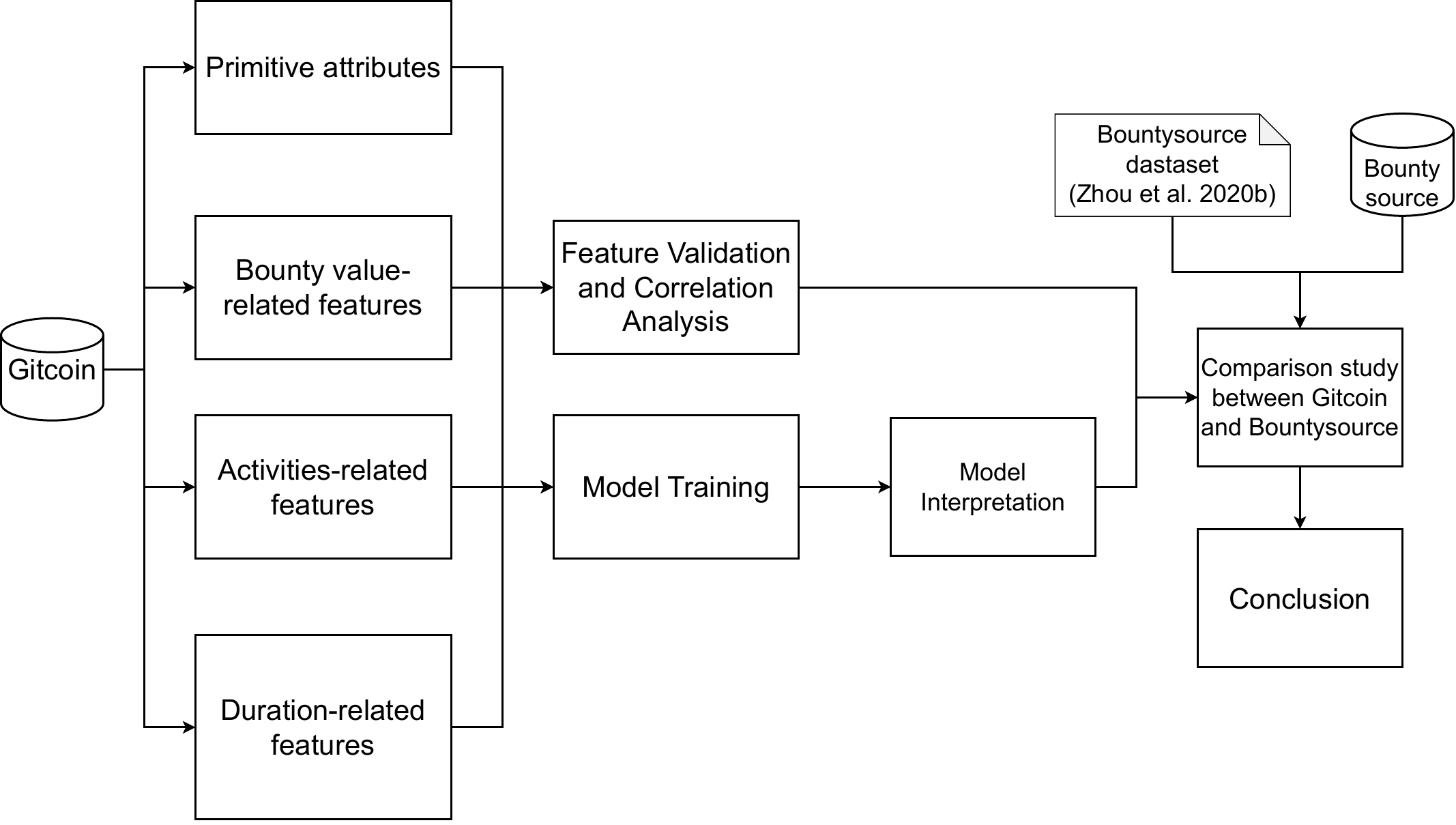}
        \caption{Overview of our study}
        \label{figure-research_design}
    \end{figure*}

    In this section, we describe the approach of our study, including data collection, preprocessing, and labeling. The overview of our study is shown in Figure \ref{figure-research_design}.

\subsection{Overview}
    
    We first collected bounty issues from the Gitcoin platform. We performed data preprocessing and labeling to identify issues that were successfully resolved, and the bounty rewards were claimed. We then perform feature extraction to characterize the Gitcoin issues. We extracted four groups of features from the issues, which are 1. primitive attributes, 2. bounty value-related features, 3. activity-related features, and 4. duration-related features. We then analyze the extracted features to investigate the factors that correlate and impact the outcome of Gitcoin's issues. In the first approach, we apply correlation analysis techniques, using the Spearman Rank Correlation Coefficient \citep{Liu2010}, to explore the correlation among each feature. \jssrev{To control for the risk of false positive results due to multiple comparisons, we use the Bonferroni correction in this analysis \citep{bonferroni} (see Section \ref{section-correlation}).}
    \jssrev{We then employed machine learning techniques, including Random forests \citep{Lan2019} and logistic regression \citep{harrell2015regression}, to construct classification models to study the relationship between the extracted features and the outcome of the issues (see Section \ref{section-classification}). Specifically, we aimed to identify the features that have a strong correlation with the outcomes of the issues.} We then applied the Point Biserial Correlation Coefficient \citep{Bonett2020} \jssrev{with Bonferroni adjustment \citep{bonferroni}} to observe the correlation between important features and the outcomes. These approaches complement each other and help to conclude our study. The data used in our study, including raw data of collected Gitcoin issues and extracted features and the source code used in the study, \jssrevtwo{are made publicly available at \url{https://doi.org/10.5281/zenodo.8313155}}. 
    \jssrev{Furthermore, to gain a comprehensive understanding of the differences and similarities in the issues, we conducted a comparison study between Gitcoin and Bountysource. This study is described in detail in Section \ref{section-compare}. To perform the study, we used the Bountysource dataset provided by Zhou et al. \citep{Zhou2020} as well as additional data that we collected.}
    

\subsection{Data collection and preprocessing}
\label{section-datacollection}
    We collected bounty issues that were created from September 2017 to December 2020 on the Gitcoin bounty platform via Gitcoin's Application Program Interface (API).\footnote{\url{https://docs.gitcoin.co/mk_rest_api/}} 
  \jssrev{Specifically, we used the \textit{bounties} API to retrieve a list of bounty issues. The API responds with files in JSON format that contains all the information we need for our analysis. To prepare the data for our study, we apply various techniques such as data preprocessing, filtering, and feature extraction to the raw JSON files. }
    Table \ref{table:dataset} shows the number of issues in our dataset. We collected a total of 6,638 bounty issues from Gitcoin. In Gitcoin, an issue can be created on two Ethereum networks: Mainnet and Rinkeby. The latter is used by developers for testing the platform, while the former is used for the actual transactions. Therefore, our study only uses the issues created on the Mainnet network. In total, our dataset contains 4,584 issues in Mainnet (69\% of the collected issues) as shown in Table \ref{table:dataset}. 

    In addition, our study focuses on the Gitcoin issue-resolving outcomes. The collected issues were thus classified into two classes which are \emph{success issues} and \emph{non-success issues}. A success bounty refers to an issue that is successfully solved, and a bounty reward has been paid out. On the other hand, a non-success bounty implies an unsuccessfully resolved and unpaid issue. These labels were determined based on the status of the issues and the bounty-paid status of the Gitcoin issues. Among those collected issues, 2,662 issues (58.1\%) were marked as success issues.
    
    \begin{table}[ht!]
    \caption{Dataset description}
    \label{table:dataset}
    \begin{tabular*}{\tblwidth}{@{}LLLRR@{}}
    \toprule
        \textbf{Total numbers of} & & & \textbf{Numbers}  & \textbf{Percentage (Mainnet)}\\ 
    \midrule
        Issues in Gitcoin & & & 6,638 & \\
        Issues in Mainnet network & & & 4,584 & \\ 
        ~~Closed issues & & & 3,744 & 81.7\%\\
        ~~Opened issues & & & 840 & 18.3\%\\ 
        ~~Success issues & & & 2,662 & 58.1\%\\ 
        ~~Non-success issues & & & 1,922 & 41.9\%\\ 
    \bottomrule
    \end{tabular*}
    \end{table}


    
\section{Features}
\label{section-feature}
    
    Table \ref{table:features} shows the list of extracted features. In this section, we discuss each feature in detail.

    

    
\begin{table*}
\caption{Feature descriptions}
\label{table:featuredesc}
\footnotesize
\resizebox{\textwidth}{!}{%
\begin{tabular}{@{}p{3cm}lp{6cm}p{5cm}@{}}
\toprule
\textbf{Group} & \textbf{Feature Names} & \textbf{Descriptions} & \textbf{\jssrev{Rationale}} \\ 
\midrule
1. Primitive attributes & bounty\_type & Type of bounty, e.g., Bug, Feature, and Security & \multirow[t]{6}{5cm}{\jssrevtwo{The primitive attributes provide essential information about the bounty and serve as foundational details that can significantly correlate with the outcome of the bounty by providing primitive information to contributors.}}   \\ 
                                   & project\_length & Relative length of the project, e.g., hours, days, weeks, months & \\ 
                    &           experience\_level           &   Recommended experience level & \\ 
                    & \jssrev{contribution\_type} &  Type of \jssrev{contribution}, e.g., Traditional, Contest, Cooperative & \\ 
                    &           github\_comments*           &   The number of comments in an issue & \\
                    &           description\_length*         &   Length of issue description & \\
\midrule
2. Bounty value-related     &           token\_name                 &   Type of token, e.g., ETH, GIT & \multirow[t]{7}{5cm}{\jssrevtwo{The bounty value-related features focus on the value of the bounty and its changes. These features reflect the incentives and rewards associated with the bounty, which can impact contributors' motivation and willingness to engage.}} \\
                    &           value\_in\_eth*              &   Value of the bounty in Ethereum & \\ 
                    &           value\_in\_usdt* & Approximation of value in US Dollars at bounty web3\_created timestamp & \\ 
                    &           value\_in\_usdt\_now*        &   Approximation of current value in US Dollars & \\ 
                    &           \jssrevtwo{token\_value\_in\_usd}        &   \jssrevtwo{The actual value of the token associated with the bounty in USD at the time of issue creation} & \\ 
                    &           value\_in\_token*            &   Amount of tokens rewarded for bounty & \\ 
                    &           increased\_bounty\_times      & The number of times that the bounty value is increased & \\ 
                    &           changed\_bounty\_value      & The value that the bounty has been increased from created to the latest increasing bounty & \\ 
\midrule
3. Activity-related & number\_of\_fulfillments* & The number of participants that submitted work to an issue & \multirow[t]{6}{5cm}{\jssrevtwo{The activity-related features are the values that relate to the activities that happened to the bounty by other developers besides the creator of the bounty. These features reflect the interest and effort of contributors, indicating a level of collaboration and involvement that can contribute to the success of an issue.}} \\
                    &           number\_of\_interests*                   & The number of participants that are interested in an issue & \\
                    &           number\_of\_activities                  & The number of activities that occur in an issue & \\
                    &           number\_of\_user\_in\_activities        & The number of usernames that shows in the issue's activities, including a funder. & \\
                    &           firstAct\_activity\_type               & The activity type of the first activity occurs & \\
                    &           lastAct\_activity\_type                & The activity type of the last activity happens & \\ 
\midrule
4. Duration-related &duration\_create\_to\_expire & The number of days between the creation of an issue and issue expiration date & \multirow[t]{8}{5cm}{\jssrevtwo{The duration-related features contain various time-related values associated with issues and their bounties. These features are aimed at providing an understanding of the temporal aspects of the bounty issues. They may offer insights into critical factors such as the time sensitivity of a bounty, the ability to set realistic deadlines, and the assessment of timeliness.}} \\
                    &           duration\_create\_to\_new\_bounty*       & The number of days between the creation of an issue and its first bounty   \\
                    &           duration\_create\_to\_worker\_applied   & The number of days between the creation of an issue and when participant request \\
                    &           duration\_create\_to\_start             & The number of days between the creation of an issue and when the participant starts working  \\
                    &           duration\_create\_to\_stop              & The number of days between the creation of an issue and when the participant stops working  \\
                    &           duration\_create\_to\_done              & The number of days between the creation of an issue and when the work is done         \\
                    &           duration\_create\_to\_submitted         & The number of days between the creation of an issue and when the work is submitted        \\
                    &           duration\_create\_to\_killed            & The number of days between the creation of an issue and when the bounty is killed       \\
\bottomrule
\multicolumn{4}{l}{\footnotesize{* adopted from \cite{Zhou2020} and \cite{Zhou2021}}}
\end{tabular}
}
\label{table:features}
\end{table*}


\subsection{Primitive attribute of the Gitcoin issue}

The primitive attributes indicate the basic information of an issue, e.g., bounty types, \jssrev{contribution} types, and length of projects. 

\begin{table}
    \caption{\jssrev{Bounty Types}}
    \label{table:bounty_types}
    \resizebox{0.45\textwidth}{!}{%
    \begin{tabular}{lp{5cm}r}
    \toprule
    \jssrev{Bounty Type} & \jssrev{Description} & \jssrev{Issues} \\
    \midrule 
    \jssrev{Feature} & \jssrev{Creating new features of a system.} & \jssrev{1,851} \\    
    \jssrev{Improvement} & \jssrev{Improving existing features, functions, or the system.} & \jssrev{679} \\
    \jssrev{Bug} & \jssrev{Creating a bug fix.} & \jssrev{333} \\
    \jssrev{Documentation} & \jssrev{Creating documentation for a system.} & \jssrev{224} \\
    \jssrev{Security} & \jssrev{Performing security-related activities, e.g., penetration test, system audit.} & \jssrev{70} \\
    \jssrev{Design} & \jssrev{Creating system or artistic design.} & \jssrev{46} \\
    \jssrev{Code Review} & \jssrev{Reviewing code} & \jssrev{20} \\
    \bottomrule
    \end{tabular}
    }
\end{table}

\subsubsection{Bounty types}
\label{subsub-bt-type}
The bounty type feature (\textit{bounty\_type}) explains the types of bounty provided in Gitcoin. This informs contributors what the bounty type of an issue is. There are 9 bounty types of an issue as depicted in Table~\ref{table:bounty_types}. The first type is \textit{Feature} type. We found 1,851 issues (40.4\%) from Mainnet which were assigned to this type. This is the largest type which reflects that an issue requires the contributors to develop or create new features of a system. The second most popular bounty type is \textit{Improvement} which requires improvements to existing features, functions, or the system that the issue owners own. We found 679 issues (14.8\%) assigned to an improvement type. The third type is \textit{Bug} (333 issues -- 7.3\%). The \textit{Bug} type indicates a bug-fixing task. For example, the issue \texttt{Not Able To Login Into WordPress From Metamask Mobile Browser}\footnote{\url{https://gitcoin.co/issue/metamask/metamask-mobile/2954/100026352}} is a bug type requiring someone to fix an error on login into the WordPress from the Metamask mobile browser. The next type is the \textit{Documentation}. This type has only 224 issues (4.9\% of the total number of issues). The \textit{Documentation} type could be related to information documentation such as programming guides, tutorials, descriptions of technologies, and translations.

    \begin{figure*}[h]
        \centering
        \input{Images/1-bounty-type-issuccess}
        \label{figure-bounty_type}
        \caption{Bounty types}
    \end{figure*}
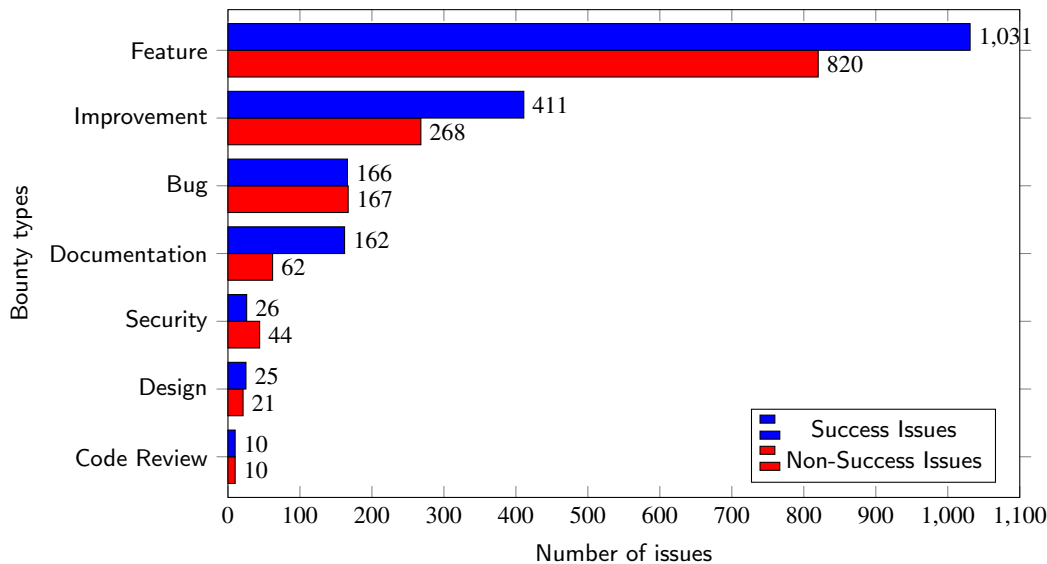    
        
    The \textit{Security} type has been found with 70 issues (1.5\%). The \textit{Security} type refers to security-related implementation tasks such as vulnerability protection, software penetration, and system audit. The next bounty type is the \textit{Design} type, which has 46 issues (1\%). The issue owners could require any design for their work in terms of flow design, wireframes, and architecture design, such as logo, schwag product, and POAP badges used for NFTs events. The last bounty issue type is the \textit{Code Review} type. We found only 20 issues (0.4\%). This bounty type requires contributors to review source code, such as defect identification, bug localization, and code quality analysis. Note that two more issue types do not have specific tasks. They are \textit{NA} type, which has 1,015 issues (22.1\%), and \textit{Other} type, which has 346 issues (7.6\%).
    
    Figure \ref{figure-bounty_type} shows the success issues of each bounty type. We found that most of the bounty types have a higher number of success issues compared to non-success issues, which are \textit{Feature} (1,031 success issues -- 55.7\%), \textit{NA} (633 success issues -- 62.4\%), \textit{Improvement} (411 success issues -- 60.5\%), \textit{Other} (198 success issues -- 57.2\%), \textit{Documentation} (162 success issues -- 72.3\%), and Design (25 success issues -- 54.4\%). In contrast, there are two bounty types (\textit{Bug} and \textit{Security}) that have a higher number of non-success issues than success issues. Note that we handled the minority of the issues (7.6\%) that were assigned to undefinable types such as \textit{0}, \textit{Andere}, and \textit{Funkcja} by grouping them into the \textit{Other} type.

\begin{table}
    \caption{\jssrev{Project lengths}}
    \label{table:project_lengths}
    \begin{tabular}{p{5cm}r}
    \toprule
    \jssrev{Length} & \jssrev{Issues} \\
    \midrule 
    \jssrev{Hours} & \jssrev{2,249} \\    
    \jssrev{Days} & \jssrev{1,164} \\
    \jssrev{Weeks} & \jssrev{276} \\
    \jssrev{Months} & \jssrev{22} \\
    \jssrev{Unknown} & \jssrev{536} \\
    \jssrev{NA} & \jssrev{337} \\
    \bottomrule
    \end{tabular}
\end{table}
    
\subsubsection{Project lengths}
\label{subsub-pj-length}
\begin{figure*}[h]
        \centering
        \input{Images/2-project-length-issuccess}
        \label{figure-project_length}
        \caption{Project lengths}
    \end{figure*}
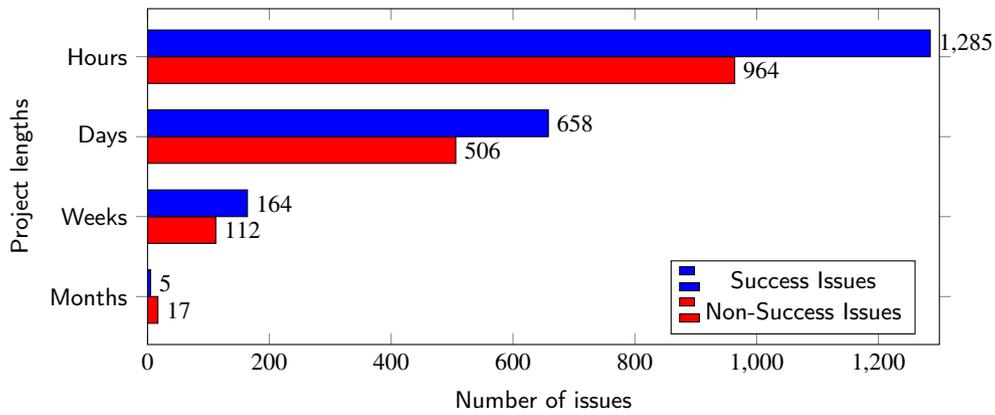

    The project length feature (\textit{project\_length}) describes the relative time duration that an issue owner expects an issue to be completed. The issue owner can select four types of project length which are \textit{Hours}, \textit{Days}, \textit{Weeks}, and \textit{Months} to provide an estimation of the task duration for contributors before committing to an issue. As shown in Table~\ref{table:project_lengths}, the top two project lengths are \textit{Hours} -- 2,249 issues (49.1\%) and \textit{Days} -- 1,164 issues (25.4\%).
    However, we found that 536 issues (11.7\%) were assigned to \textit{Unknown} and 337 issues (7.4\%) were assigned to \textit{NA}. According to Figure \ref{figure-project_length}, most issues in the project lengths of Hours, Days, and Weeks are success issues. However, the project lengths of Months have 70\% of issues that are non-success.

\begin{table}
    \caption{\jssrev{Experience levels}}
    \label{table:exp_levels}
    \begin{tabular}{p{5cm}r}
    \toprule
    \jssrev{Exp. Level} & \jssrev{Issues} \\
    \midrule 
    \jssrev{Beginner} & \jssrev{2,415} \\    
    \jssrev{Intermediate} & \jssrev{859} \\
    \jssrev{Advanced} & \jssrev{515} \\
    \bottomrule
    \end{tabular}
\end{table}
    
\subsubsection{Experience levels}
\label{subsub-exp-level}
    The required experience level feature (\textit{experience\_level}) is a feature that declares the required experience level of contributors who potentially participate in an issue. An issue owner can specify the experience level to ensure that a contributor can resolve an issue. In Gitcoin, we found three experience level values, including \textit{Beginner}, \textit{Intermediate}, and \textit{Advanced}. We can see from Table~\ref{table:exp_levels} and Figure \ref{figure-experience_level} that the top required experience level in Gitcoin is the \textit{Intermediate} level (2,415 issues -- 52.7\%)\footnote{Note that we also found the experience level named in German (\emph{Mittlere}) and Polish (\emph{Pośredni}) which means `Intermediate' in English. Thus, we then grouped this data into an Intermediate level as well.}. It also shows that the \textit{Intermediate} level has a large number of success issues (58.2\%) followed by the \textit{Beginner} level (18.7\%). Interestingly, 2 experience levels have a higher non-success rate than the success ones: \textit{Advanced} and the \textit{Other} levels.

    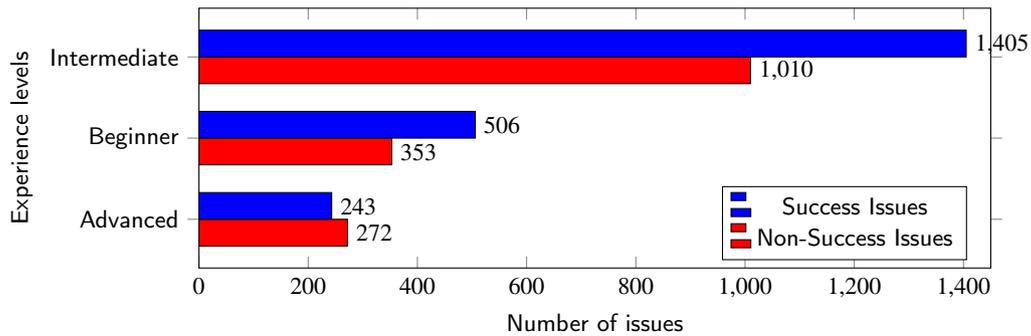
\begin{figure*}[h]
        \centering
        \input{Images/3-exp_level_issuccess}
        \caption{Experience levels}
        \label{figure-experience_level}
    \end{figure*}
        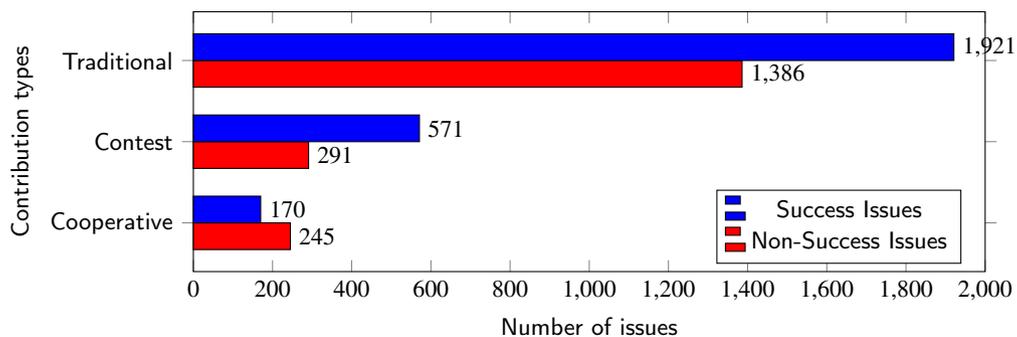
\begin{figure*}[h]
        \centering
        \input{Images/4-project_types_issuccess}
        \caption{\jssrev{Contribution} types}
        \label{figure-project_type}
    \end{figure*}
    \begin{figure*}[h]
        \centering
        \input{Images/9-github-comments-issuccess}
        \label{figure-github_comments}
        \caption{GitHub comments}
    \end{figure*}
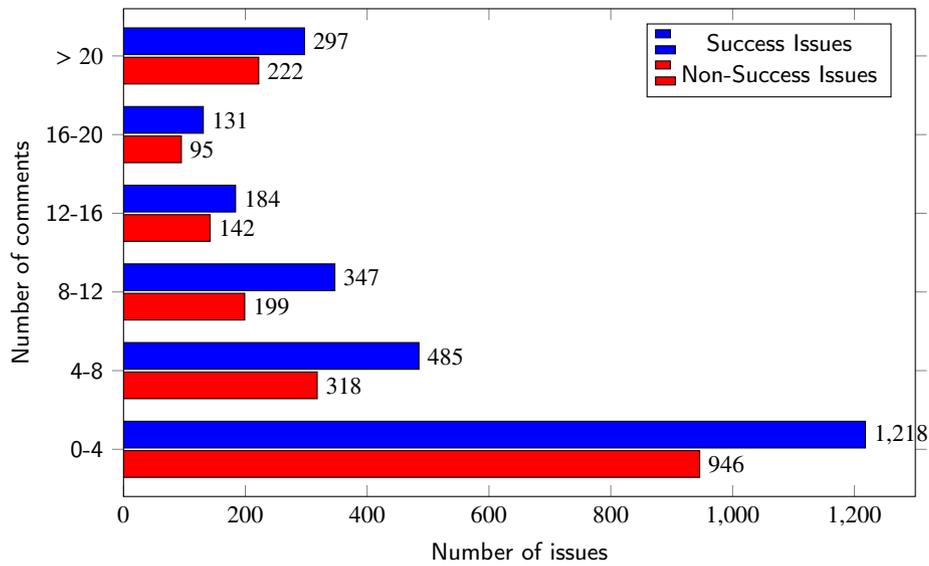
        \begin{figure*}[h]
        \centering
        \input{Images/10-desc-length-issuccess}
        \caption{Length of description}
        \label{figure-description_length}
    \end{figure*}
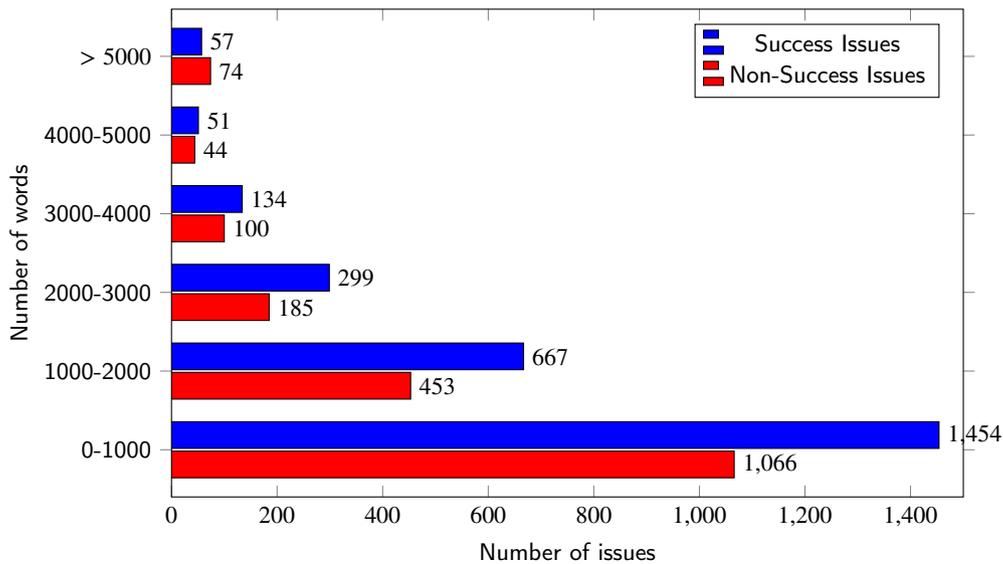   

\subsubsection{\jssrev{Contribution types}}
    The \jssrev{contribution} type feature (\textit{\jssrev{contribution\_type}}) refers to the types of \jssrev{contribution} provided in Gitcoin: \textit{Tradition}, \textit{Contest}, and \textit{Cooperative}. The \textit{Tradition} type means that there is only one contributor who can be approved to contribute and get a bounty reward. This is the most popular \jssrev{contribution} type containing 3,307 issues (72.1\%). In contrast, the \textit{Contest} type allows a number of contributors to work on an issue, but only one can be paid. There are 862 issues (18.8\%) of the \textit{Contest} type found. The least popular one is the \textit{Cooperative} type, which contains 415 issues (9.1\%). It allows several contributors to work on the issue, and the issue owners can decide to pay the bounty to more than one contributor.

    Figure \ref{figure-project_type} shows that the \textit{Tradition} type has the highest number of success issues (1,921 success issues -- 58.1\%) and also has a higher number of success issues than non-success issues. Similarly, the \textit{Contest} type has a higher number of success issues than non-success issues (571 success issues -- 66.2\%). However, the Cooperative type is the only one with a higher number of non-success issues than its counterparts (245 non-success issues -- 59\%).

\subsubsection{GitHub comments}
\label{subsub-github-comments}
    The GitHub comment feature (\textit{github\_comments}) indicates the number of comments from a corresponding GitHub issue. The GitHub issue tracking system is a platform used for tracking the contributors' work and addressing an issue they are working on. It can also be used as an intermediary channel of communication between issue owners and contributors. Comments can be the issue description, work discussion, or additional information on an issue. Figure \ref{figure-github_comments} shows the number of comments posted on the GitHub issues. It shows that the maximum number of comments reaches over 25 comments, and the minimum is 0. However, we found that the issues in any number of comments have a higher success rate than the non-success rate with 6 comments on average.
    

\subsubsection{Length of description}
\label{subsub-description-length}
    The length of the description feature (\textit{description\_length}) describes the relative length of the issue description (i.e., number of characters). This feature is extracted to observe whether the description length correlates with the outcome of an issue. Figure \ref{figure-description_length} shows the length of the issue description from the collected issues. We have noticed that the numbers of success issues and non-success issues are almost equal in those issues that have a longer description rather than those shorter description issues which have a higher number of success issues. 
    
    
    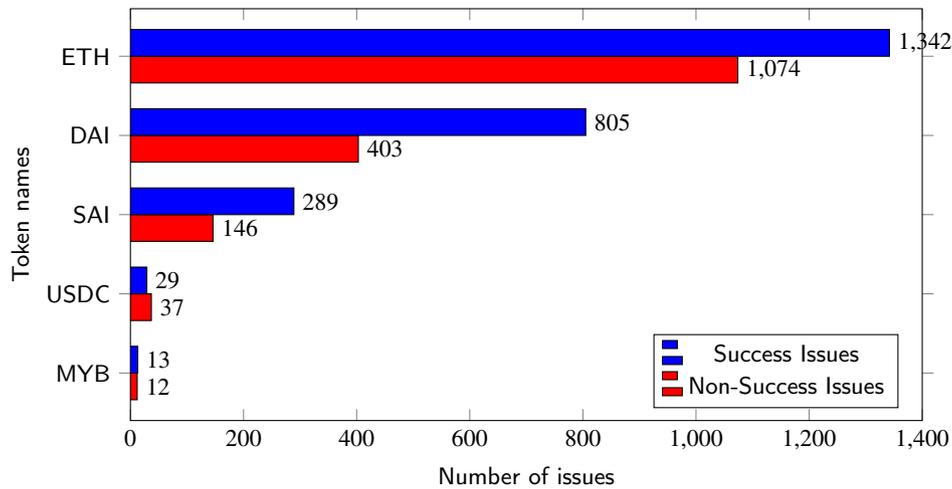
\begin{figure*}[h]
        \centering
        \input{Images/5-token-names-issuccess}
        \caption{Token names}
        \label{figure-token_name}
    \end{figure*}

\subsection{Bounty related features}
The bounty value-related features reflect the value of the bounty reward of an issue, e.g., value in ETH, increasing bounty value, and token names.

\subsubsection{Token names}
    The token names feature (\textit{token\_name}) indicates the types of cryptocurrency tokens provided as a bounty reward given to the contributors who can resolve bounty issues in Gitcoin. Nevertheless, over one hundred token types are found on the platform, and we then picked the 5 most popular token types (Figure \ref{figure-token_name}). The \textit{Ethereum (ETH)} is the most popular token type that has been used in Gitcoin projects with 2,416 issues, accounting for 52.7\% of all the issues. The \textit{Dai} token (DAI) is the second most popular token, with 1,208 issues (26.4\%). The second token is the \textit{Sai} token (SAI) found 435 issues (9.5\%), followed by \textit{USD} Coin token (USDC) which consists of 66 issues (1.4\%). The last one is the \textit{MyBit} token (MYB), which has 25 issues or only 0.6\% of the total issues. According to the token types, most of them are operated under the Ethereum blockchain, as explained in Section~\ref{subsection-gitcoin}. We found that the USDC token has 56.1\% for non-success issues in the platform.


\subsubsection{Bounty value}
\label{subsub-bValue}
	    
    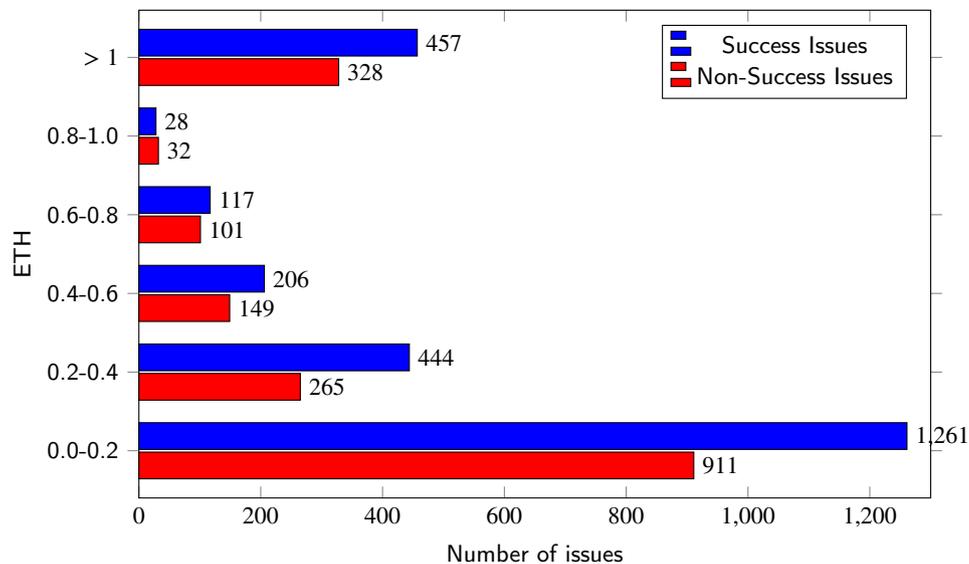
\begin{figure*}[h]
        \centering
        \input{Images/11-value_in_eth-issuccess}
        \label{figure-value_in_eth}
        \caption{ETH bounty value}
    \end{figure*}

    \begin{figure*}[h]
        \centering
        \input{Images/12-value_in_usdt-issuccess}
        \caption{USDT bounty value}
        \label{figure-value_in_usdt}
    \end{figure*}
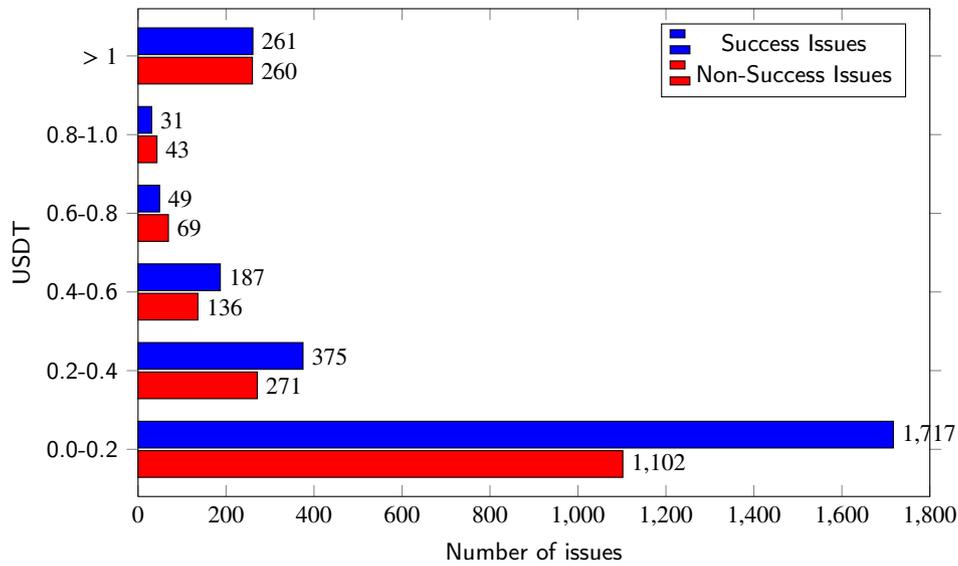

    \begin{figure*}[h]
        \centering
        \includegraphics[width=0.8\textwidth]{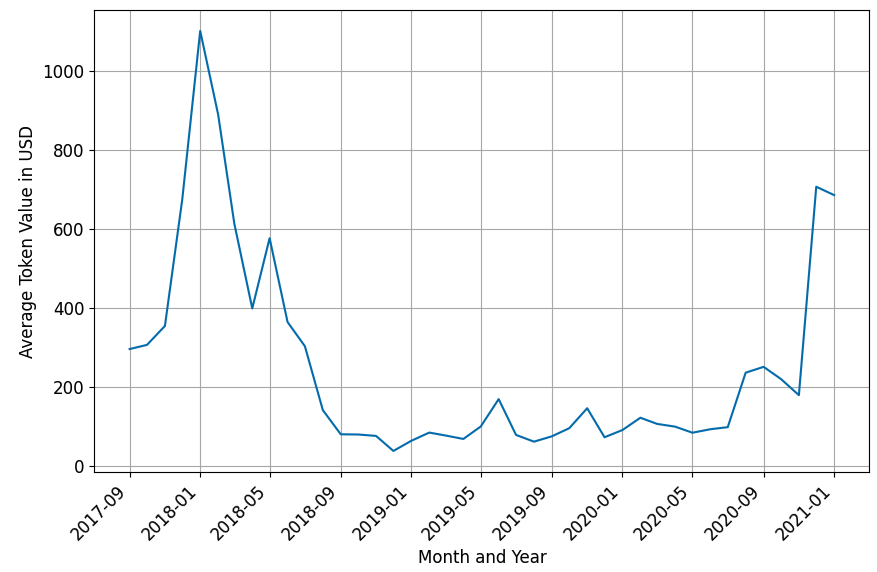}
        \caption{Average token value in USD over time across all token names}
        \label{figure-tokenValueinUSD}
    \end{figure*}
     
    For the features that are related to the bounty value in Gitcoin, we particularly focus on the bounty value in Ether or ETH (\textit{value\_in\_eth}) since it is a token worked under the Ethereum blockchain and is mostly used as a bounty reward in the Gitcoin platform. Figure \ref{figure-value_in_eth} shows the success of the issues with ETH tokens. The maximum value of the ETH that has been proposed is 1.14 (0.2 on average). Additionally, we found that the issues are mostly successful in the range of bounty values between 0.1 to 0.4 ETH. On the other hand, the upper values have a higher non-success rate than the success rate. Since the ETH tokens are in digital currency, we also observe those values in U.S. Dollars (USD). Therefore, we extracted the Tether or USDT value (\textit{value\_in\_usdt}) feature from Gitcoin issues to refer to the bounty value in USD since the USDT can imply a stablecoin of the U.S. Dollars \citep{Grobys2021}. Figure \ref{figure-value_in_usdt} shows the success of the issues with USDT token, in which we found that those values corresponded to the values in ETH. \jssrevtwo{Moreover, Figure \ref{figure-tokenValueinUSD} shows the average token value in USD (\textit{token\_value\_in\_usd}) over time across all token names. It reflects the market values of various cryptocurrencies. It is observed that the majority of tokens experienced a decline in value during the year 2019. This downturn was followed by a notable recovery and upward trend in token values during the year 2021.}

    

\subsubsection{The changes of bounty values}
    The issue owners can determine whether to increase or decrease a bounty value for their issues, such as attracting contributors to work on issues. We extract two features related to the changes in bounty values: the number of times that bounty has been increased (\textit{increased\_bounty\_times}) and the total bounty value that has been changed (\textit{changed\_ bounty\_value}). The former explains how frequently an issue has been increased by an issue owner. The latter is calculated based on the changed bounty value based on the difference between the most recent value and the original value, i.e., the value at the first time that bounty has been proposed. This feature is to observe how much issue owners have to increase their bounty values to attract contributors. We found that 63\% of those issues having their bounty value increased were success issues. However, the bounty values have never been changed in most of the issues.
    
    

\subsection{Activity related features}
\label{act-related}

The activity-related features indicate activities that occur with an issue, e.g., the number of interests and the number of activities.

\subsubsection{Number of fulfillments}
    The number of fulfillments (\textit{number\_of\_fulfillments}) indicates the number of participants who submitted the work to issue owners.  We found that 67.8\% of total issues contain at least one fulfillment, and the highest number is 156 participants. Moreover, the number of issues with a large number of fulfillments also has higher success issues.

\subsubsection{Number of interests}
    The number of interests (\textit{number\_of\_interests}) indicates the number of participants who are interested in working on issues. From our investigation, only 19.3\% of issues have zero interests, while 3,689 issues (80.7\%) have at least one interest from contributors.
    
\subsubsection{Number of activities}
    The number of activities (\textit{number\_of\_activities}) indicates the total number of activities that occurred in an issue. We extracted the activities from Gitcoin's issue change log. Those activities include proposing bounties, increasing bounty values, approving candidates, and making submissions. 
    
    
\subsubsection{Number of users who interact with an issue}
    This feature (\textit{number\_of\_user\_in\_activities}) is the total number of users who perform actions on an issue. On average, there are 5 users who interact with an issue. The highest number of users that interact with one issue is 170 users.

\subsubsection{The first activity type occurred on an issue}
    This feature \textit{firstAct\_activity\_type} indicates the activity type that occurred in an issue after issue creation. We found that proposing bounty, start working, and worker applied are the top three types that appeared as the first activity. 
    
\subsubsection{The last activity type occurred on an issue}

    This feature \textit{lastAct\_activity\_type} captures the last activity type that occurred on an issue. Although we found that the majority type is the work submission activity (\textit{work\_submitted}) which indicates the resolving of issues, we have noted that the bounty may not be paid, which causes a non-success issue. We found that over 300 issues have been marked as submitted, but the bounties were not paid.




\subsection{Duration-related features} 

    We determine the duration from the issue creation time to each stage of issues (i.e., issue status) to investigate the relationships between durations and the issue-addressing outcome. We extract eight duration-related features.  
    
    \begin{figure*}[h]
        \centering
        \input{Images/7-all-duration}
        \caption{The duration from issue creation to each stage} 
        \label{figureallduration}
    \end{figure*}
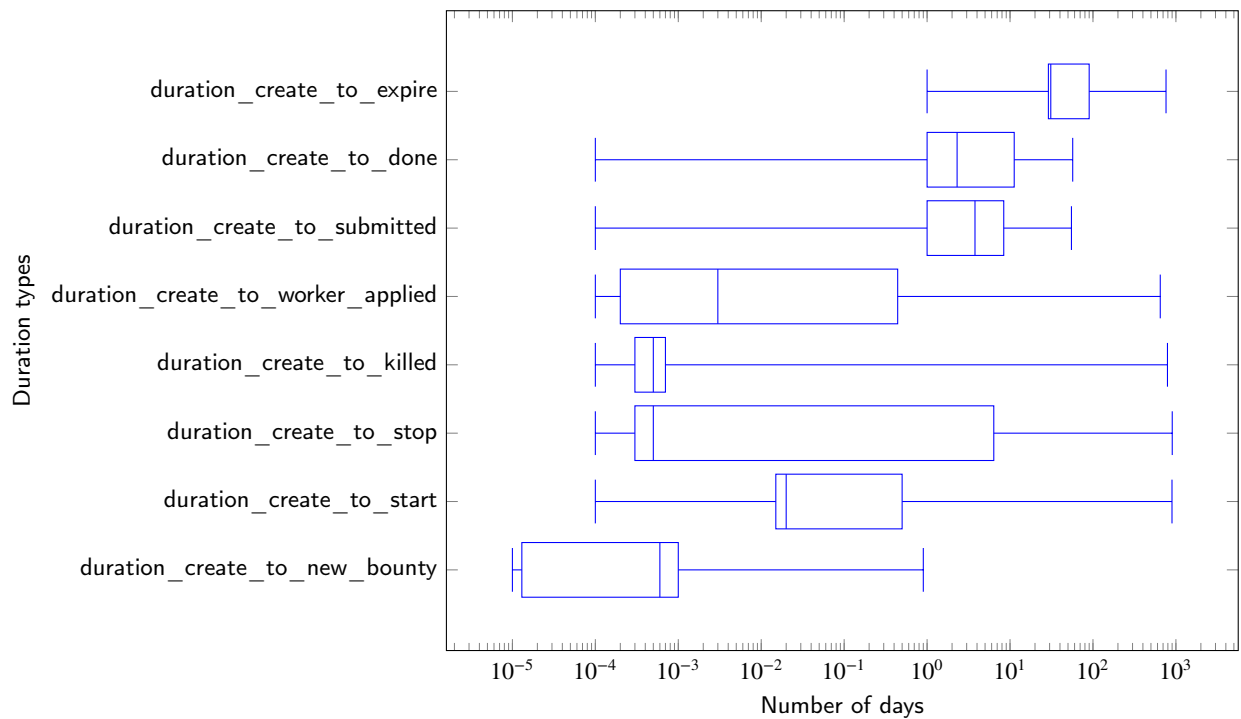

    Figure \ref{figureallduration} shows the extracted duration-related features. We found that issue owners usually add bounty rewards 10 minutes after the issue creation (\textit{duration\_create\_to\_new\_bounty}). In addition, contributors mostly apply to work on an issue on day six after issues were created to apply to issues (\textit{duration\_create\_to\_worker\_applied}) and spent, on average, eight days to resolve issues.

\section{Feature and correlation analysis} 
\label{section-correlation}
    
    \begin{figure*}[h]
        \centering
        \input{Images/bounty-types-exp-levels}
        \caption{The number of issues in each experience level and each bounty type  
        }
        \label{figure-analyze1}
    \end{figure*}
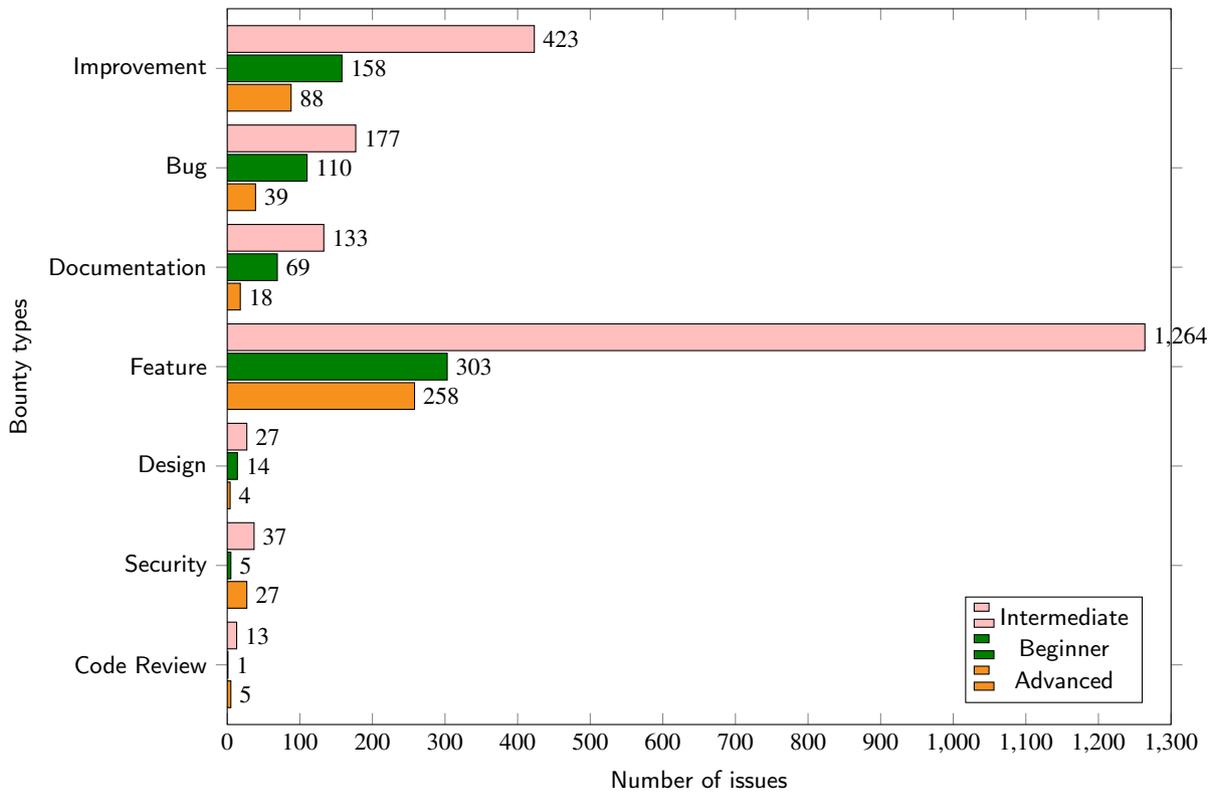
    
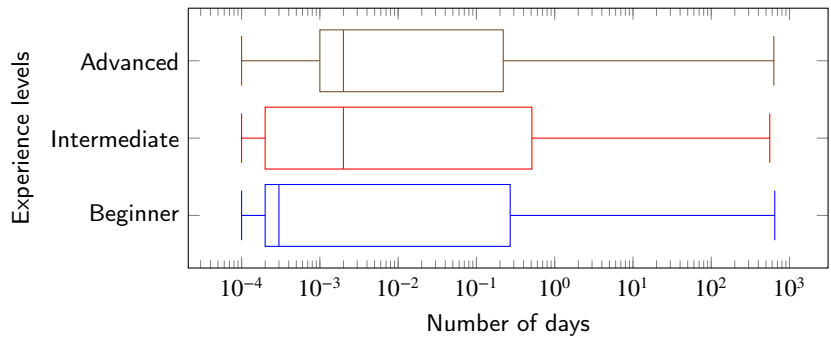
\begin{figure*}[h]
        \centering
        \input{Images/exp-level-worker-applied}
        \caption{The duration to find a contributor}
        \label{figure-analyze3}
    \end{figure*}

    Figure \ref{figure-analyze1} shows the number of issues in each experience level and each bounty type. We found that the number of issues that are required the experience level of \textit{Intermediate} is the majority in several bounty types such as Improvement, Bug, and Feature. Therefore, this experience level is appropriate for various contributors. In addition, we found that the intermediate-level issues mostly relate to language translation, data migration/integration, and developing features in open-source projects. For the \textit{Advanced} level, the issues are mostly related to complex development tasks which require contributors that are proficient in applying technical knowledge since they need to highly understand the complication of work to implement the work successfully. In addition, we investigate the relationship between the different experience levels and their duration. We found that, on average,  the \textit{Intermediate} level issues took five days to find a contributor while the other levels took six days (Figure \ref{figure-analyze3}). The resolution duration of the majority of issues at all levels is between six to eight days. 
    
    
    
    
We then apply the Spearman Rank Correlation technique \citep{Liu2010} to determine the correlation among the features. \jssrev{To account for multiple comparisons, we adjust the significance level using the Bonferroni correction. In this study, our focus is to investigate the correlation between each pair of features. The null hypothesis to be rejected is that there is no correlation between the two features being compared. If the adjusted p-value is less than the significance level, we reject the null hypothesis and conclude that there is a statistically significant correlation between the two features. The pairs of features with a strong correlation and a statistically significant p-value \jssrevtwo{($p < 0.05/\text{number of comparisons}$)} are reported as follows.}


\textbf{Positive correlation between Project length and Experience level}

     The correlation between \textit{Project length} and \textit{Experience level} is a moderately positive correlation \jssrevtwo{($p < 0.05/$\text{number} \text{of comparisons}$)$}. From our investigation, we found the majority of issues that requires the beginner and intermediate experience levels are expected to be resolved within hours and days, while the issues that require the advanced level are mostly expected to be resolved in the length of months. 

\textbf{Positive correlation between the number of times that bounty has been increased and the total bounty value that has been changed}
    The number of times that bounty has been increased and the total bounty value that has been changed is positively correlated. This indicates that the higher number of times that issue owner increases bounty, the higher the bounty value added.
    
\section{\jssrevtwo{Identifying features correlated with the outcome of the Gitcoin bounty issue}}
\label{section-classification}

\jssrev{To identify a set of features that strongly correlate with the outcome of the Gitcoin bounty issue, we adopt two machine learning techniques to train classifiers to determine whether an issue will be a success or a non-success issue. Firstly, we adopt Random Forests (RF) \citep{BREIMAN2001} to train a classifier. By using RF, we can also identify feature importance which can determine the strength of association between features and the target variable (i.e., successful issue resolution). RF uses the \textit{Gini} impurity measure to calculate the feature importance to explain to determine the significance of each feature in determining the output. Since RF provides an interpretable classification in terms of the impact of each feature on the outcome, this method has been applied in various empirical studies, e.g., \cite{Laaber2021,Tagra2022}. In addition to Random Forests, this study employs Logistic Regression Modeling (LG) and removing multicollinearity \citep{harrell2015regression}, as seen in previous empirical studies in software engineering, e.g., \cite{Thongtanunam2017}. In LG, a model derives a set of coefficients that represent the relationship between each feature and the binary outcomes (i.e., success or non-success) The coefficients can be interpreted as the change in the log-odds of the outcome as in a change of the feature variable. In terms of interpretability, we then apply the Wald test statistic to measure the strength and direction of the relationship between each feature and the outcomes.}

    
    We split the dataset into two parts: 70\% for the training set and 30\% for the test set. We also applied the stratified sampling technique \citep{Ye2013} to preserve the proportion of the target variable in the training and test data set and applied the bootstrap sampling technique to overcome the over-fitting problem. Table \ref{table:traintest} shows the number of issues in our dataset. In our study, we conduct two different experiments that aim to identify the importance of features in two scenarios by varying a set of features that are used in model training. 
    
    In the first experiment (Setting 1), we used all extracted features (Table \ref{table:features}) in model training to identify important features among all features that characterized the bounty issues. Since we also aim to provide suggestions to the issue owners when they create a bounty issue. In the second experiment (Setting 2), we thus used only the features that can be manipulated by the issue owners at their issue creation time. We can then identify a set of important features at the issue creation time. We then discuss the top important features. Note that we replace the values of \textit{project\_length} and \textit{experience\_level} with the ordinal values to reflect their meaning. For example, the \textit{experience\_level} was mapped as follows: 1 for \textit{Beginner}, 2 for \textit{Intermediate}, and 3 for \textit{Advanced}. 
    

    \begin{table}[h!]
    \caption{Experimental setting}
    \centering
    \begin{tabular*}{\tblwidth}{@{}l r r@{}}
        \toprule
        & \myindent \myindent \myindent \textbf{Success Issue} & \textbf{Non-success Issue} \\
        \midrule
            \multirow{1}{*}{Training Set} & 1,861 & 1,348 \\
            \multirow{1}{*}{Test Set} & 797 & 578 \\
        \bottomrule
    \end{tabular*}
    \label{table:traintest}
    \end{table}

    \begin{table}[h!]
    \caption{\jssrev{\jssrevtwo{Evaluation Results of each experimental setting using Random Forests (RF),  Logistic Regression (LG), and Baseline}}}
    \centering
    \begin{tabular*}{\tblwidth}{@{}l rrrr@{}}
        \toprule
        \textbf{} & \myindent \textbf{Precision} & \textbf{Recall} & \textbf{F1} &\textbf{Acc.} \\
        \midrule
            \multirow{1}{*}{RF:Setting 1} & 0.99 & \jssrevtwo{0.98} & \jssrevtwo{0.99} & \jssrevtwo{0.99} \\
            \multirow{1}{*}{RF:Setting 2} & 0.85 & 0.91 & 0.88 & 0.86 \\\midrule
            
\multirow{1}{*}{\jssrev{LG:Setting 1}} & \jssrevtwo{0.77} & \jssrevtwo{0.77} & \jssrevtwo{0.77} & \jssrevtwo{0.77} \\
            \multirow{1}{*}{\jssrev{LG:Setting 2}} & \jssrev{0.70} & \jssrev{0.70} &\jssrev{0.70} &\jssrev{0.70} \\   
            \midrule
            \multirow{1}{*}{\jssrevtwo{Baseline}} & \jssrevtwo{0.51} & \jssrevtwo{0.50} &\jssrevtwo{0.50} &\jssrevtwo{0.58} \\   
        \bottomrule
    \end{tabular*}
    \label{table:experimental}
    \end{table}
    
    Table \ref{table:experimental} shows the performance evaluation in terms of precision, recall, F1-score, and accuracy of the models trained from the two settings \jssrev{using Random Forest (RF) and Logistic Regression (LG)}. \jssrev{The results show that, in both settings from the two models, it can accurately classify the issue outcomes by achieving over 70\% in all measurements.} 
    \jssrevtwo{We however acknowledge the lower performance of logistic regression (LG) compared to Random Forests (RF). However, it is important to consider that LG and RF offer different advantages in terms of model interpretability and feature importance identification. While the importance of features in a Random Forest model helps identify which features have the most predictive power, LG models provide greater interpretability due to their linear nature. The coefficients in LG allow us to analyze the impact of each feature on the output, providing valuable insights into the relationship between features and the target variable. By combining insights from both models, we can achieve a more comprehensive understanding of the data and uncover underlying patterns. Moreover, both RF and LG also identify a common set of important features such as project length and experience level. In addition, to address concerns about the models' acceptance, we have included a baseline method using Zero-Rule as a sanity check. Zero-Rule predicts the most frequent label in the training set, serving as a simplistic benchmark. Both LG and RF outperform this baseline method, indicating their superiority in predictive performance. Therefore, it is reasonable to use these models for our interpretation and analysis. By considering the interpretability strengths of LG and the feature importance of RF, and validating their performance against a baseline, we ensure a robust evaluation of the models' effectiveness and reliability for our study.}

    \subsection{Analysis of the feature importance from Random forests}
    
    Table \ref{table:importance1} shows the list of the top ten important features from the first and second settings, respectively. The top three highest important features of the first setting are  \textit{duration\_create\_to\_done}, \textit{duration\_create\_to\_submitted}, and \textit{lastAct\_activity\_type\_work \_done}. As can be seen that those three most important features reflect a common scenario in software development that a proper time duration spent on an issue is critical to the success of issue resolution. \jssrev{However, it is worth noting that these features can only be gathered when the issues are closed.}

    In the second setting, we thus use the features that can be manipulated or controlled by the issue owner, \textit{description\_length}, \textit{duration\_create\_to\_new\_bounty}, and \textit{value\_ in\_usdt} are the top three most important features which must be taken into account when creating issues. In particular, we found that the number of days from issue creation until the first proposed bounty is the second most important feature that potentially determines the issue outcome. This finding corresponds with the result reported in \citep{Zhou2020}, which also found that the earlier bounty proposed, the higher likelihood of being addressed. We then further investigate the correlation between these features.

   \begin{table}[h!]
    \caption{\jssrev{Feature importance from the two experimental settings using Random Forests}}
    \centering
    \begin{tabular*}{\tblwidth}{@{}lr@{}}
        \toprule
        \textbf{Features} & \textbf{Importance Values} \\
        \midrule
        \multirow{1}{*}{\jssrev{\textbf{Setting 1:}}} & \\
        \multirow{1}{*}{duration\_create\_to\_done} & \jssrevtwo{0.229}\\
        \multirow{1}{*}{duration\_create\_to\_submitted} & \jssrevtwo{0.076} \\
        \multirow{1}{*}{lastAct\_activity\_type\_work\_done} & \jssrevtwo{0.062}\\
        \multirow{1}{*}{number\_of\_fulfillments} & \jssrevtwo{0.048} \\
        \multirow{1}{*}{duration\_create\_to\_killed} & \jssrevtwo{0.044} \\
        \multirow{1}{*}{\jssrevtwo{token\_value\_in\_usdt}} & \jssrevtwo{0.042} \\
        \multirow{1}{*}{number\_of\_activities} & \jssrevtwo{0.035} \\
        \multirow{1}{*}{lastAct\_activity\_type\_killed\_bounty} & \jssrevtwo{0.034}\\
        \multirow{1}{*}{number\_of\_interests} & \jssrevtwo{0.034} \\

        \midrule
         \multirow{1}{*}{\jssrev{\textbf{Setting 2:}}} & \\
        \multirow{1}{*}{description\_length } & 0.143 \\
        \multirow{1}{*}{duration\_create\_to\_new\_bounty} & 0.139 \\
        \multirow{1}{*}{value\_in\_usdt} & 0.126 \\
        \multirow{1}{*}{value\_in\_usdt\_now} & 0.102 \\
        \multirow{1}{*}{duration\_create\_to\_expire} & 0.097 \\
        \multirow{1}{*}{value\_in\_eth} & 0.089 \\
        \multirow{1}{*}{value\_in\_token} & 0.052 \\
        \multirow{1}{*}{experience\_level\_code} & 0.037 \\
        \multirow{1}{*}{project\_length\_code} & 0.037 \\
        \multirow{1}{*}{bounty\_type\_Feature} & 0.016 \\
        \bottomrule
    \end{tabular*}
    \label{table:importance1}
    \end{table}

    \jssrev{We apply the Point Biserial Correlation Coefficient \citep{Bonett2020} with Bonferroni correction to measure the correlation between the numerical features and the binary outcomes of issues (i.e., success and non-success issues). Table \ref{table:pointbiserial} lists the features used in Setting 2 along with their correlation coefficient and p-value. Significant correlations at a statistically strong level \jssrevtwo{($p < 0.05 / \text{number of features}$)} are indicated in the table. We used the features in Setting 2 for our analysis because they are manipulable and more useful to practitioners.}
    
\begin{table}[]
\centering
\caption{\jssrev{The correlation of the features with the issue outcomes using the Point Biserial Correlation Coefficient}}
\label{table:pointbiserial}
\resizebox{0.45\textwidth}{!}{%
\begin{tabular}{@{}lrr@{}}
\toprule
\jssrev{Features}                    & \multicolumn{1}{l}{\jssrev{Correlation}} & \multicolumn{1}{l}{\jssrev{p-value}} \\ \midrule
\jssrev{token\_name\_DAI}            & \jssrev{0.110} $\uparrow$                          & \jssrevtwo{$<$ 0.001}                     \\
\jssrev{bounty\_type\_Documentation} & \jssrev{0.097} $\uparrow$                          & \jssrevtwo{$<$ 0.001}                     \\
\jssrev{contribution\_type\_contest}      & \jssrev{0.062} $\uparrow$                        & \jssrevtwo{$<$ 0.001}                     \\
\jssrev{token\_name\_SAI}            & \jssrev{0.055} $\uparrow$                         & \jssrevtwo{$<$ 0.001}                     \\
\jssrev{bounty\_type\_NA}            & \jssrev{0.049} $\uparrow$                            & \jssrev{ 0.001}                       \\
\jssrev{bounty\_type\_Security}      & \jssrev{-0.070} $\downarrow$                       & \jssrevtwo{$<$ 0.001}                     \\
\jssrev{experience\_level\_code}     & \jssrev{-0.070} $\downarrow$                         & \jssrevtwo{$<$ 0.001}                     \\
\jssrev{token\_name\_ETC}            & \jssrev{-0.075} $\downarrow$                         & \jssrevtwo{$<$ 0.001}                     \\
\jssrev{token\_name\_DOT}            & \jssrev{-0.089} $\downarrow$                        & \jssrevtwo{$<$ 0.001}                     \\
\jssrev{contribution\_type\_cooperative}  & \jssrev{-0.110} $\downarrow$                        & \jssrevtwo{$<$ 0.001}                     \\ 
\bottomrule
\end{tabular}%
}
\end{table}

\jssrev{The studying result suggests that the features ``token\_name \_DAI'' and ``bounty\_type\_Documentation'' have the strongest positive correlation with issue outcomes, while the features ``contribution\_type\_cooperative'' and ``token\_name\_DOT'' have the strongest negative correlation with issue outcomes. The results are statistically significant, with all p-values falling below the significance threshold of $0.05$ after the Bonferroni correction. It shows that the type of token used and the type of bounty offered could be strongly correlated with issue success. The results suggest that the token ``DAI'' is positively correlated with issue success, potentially due to the stability of the coin. Regarding the contribution types, ``Documentation'' bounties are positively correlated with issue success, while ``Security'' bounties are negatively correlated. This may reflect the fact that documentation bounties are easier to define and assess, while security bounties require more expertise to evaluate and address. }

\subsection{\jssrev{Analysis of the feature's coefficients from Logistics regression}}

\begin{table}[h]
\centering
\caption{Coefficient and p-value estimates from logistic regression analysis of issue success.}
\label{tab:coefficient_LG}
\resizebox{0.4\textwidth}{!}{%
\begin{tabular}{lrr}
\toprule
Feature & \multicolumn{1}{l}{Coefficient} & \multicolumn{1}{l}{p-value} \\ 
\midrule
token\_name\_DAI & 0.619 $\uparrow$ & $<$ 0.001 \\
token\_name\_ETH & 0.130 $\uparrow$ & 0.015 \\
project\_length\_code & -0.008 $\downarrow$ & 0.015 \\
bounty\_type\_Feature & -0.039 $\downarrow$ & 0.589 \\
bounty\_type\_NA & -0.068 $\downarrow$ & 0.582 \\
contribution\_type\_traditional & -0.132 $\downarrow$ & 0.009 \\
experience\_level\_code & -0.159 $\downarrow$ & $<$ 0.001 \\ 
\bottomrule
\end{tabular}%
}
\end{table}

\jssrev{Table \ref{tab:coefficient_LG} presents the results of a logistic regression analysis of issue success. The coefficients of each feature indicate the direction of the correlation with issue success, while the p-values provide information on the statistical significance of each correlation. The analysis shows that issues associated with the tokens ``DAI'' and ``ETH'' are more likely to be successful. The strong correlation between ``token\_name\_DAI'' and issue success is further supported by the results of the point biserial correlation analysis. Additionally, shorter project length, traditional contribution type, and lower experience level are all negatively correlated with issue success. This suggests that shorter projects may increase the chance of issue success, and the issues that require less experienced contributors may be more successful. Furthermore, the contest contribution type is preferred over cooperative and traditional. }

\subsection{The comparison study between Gitcoin and Bountysource}
\label{section-compare}

\jssrev{This study compares some characteristics of bounty issues from Gitcoin and Bountysource, with the aim of providing useful recommendations to bounty issue funders and contributors on selecting the most appropriate platform to meet their needs. Specifically, we examine three aspects of these platforms: the programming languages used, the topics of issues posted, and the value of bounties offered. To conduct this study, we utilized the Bountysource dataset provided by \cite{Zhou2020}, which includes valuable information on bounty issues in the Bountysource platform. In addition, we collected supplementary data, such as the topics of each issue corresponding to the issue key provided in the dataset.}

\jssrev{Our first comparison is to investigate the programming languages used in bounty issues across both platforms. To identify the programming languages used in bounty issues across both platforms, we collected language-related tags from the GitHub repositories associated with each bounty issue in both datasets. Table \ref{tab:language} shows the percentage distribution of programming languages used in bounty issues on both Gitcoin and Bountysource platforms. The results show the differences in the distribution of programming languages between the two platforms. JavaScript is the most commonly used language on both platforms, with a distribution of 43.90\% on Gitcoin and 17.50\% on Bountysource. This could be attributed to the popularity of JavaScript in web development, which may be a primary focus of the bounty issues on both platforms. In addition, Bountysource has a higher percentage of C++ and Python, which is also widely used in general software development. However, the difference in language distribution may also reflect the types of projects and bounties available on each platform. Gitcoin, for example, has a focus on blockchain applications, which may explain the higher percentage of TypeScript and Solidity, which are both used in the development of smart contracts and other blockchain-related applications.}

\begin{table}[]
\centering
\caption{\jssrev{Percentage distribution of programming languages in Bountysource and Gitcoin}}
\label{tab:language}
\resizebox{0.32\textwidth}{!}{%
\begin{tabular}{@{}lrlr@{}}
\toprule
\jssrev{Bountysource} & \multicolumn{1}{c}{\jssrev{\%}} & \jssrev{Gitcoin}    & \multicolumn{1}{c}{\jssrev{\%}} \\ \midrule
\jssrev{JavaScript}   & \jssrev{17.50}                 & \jssrev{JavaScript} & \jssrev{43.90}                 \\
\jssrev{C++}          & \jssrev{14.20}                 & \jssrev{TypeScript} & \jssrev{12.50}                 \\
\jssrev{Python}       & \jssrev{12.40}                 & \jssrev{Go}         & \jssrev{9.50}                  \\
\jssrev{PHP}          & \jssrev{9.10}                  & \jssrev{Python}     & \jssrev{7.70}                  \\
\jssrev{C\#}           & \jssrev{7.60}                  & \jssrev{Clojure}    & \jssrev{4.00}                  \\
\jssrev{C}            & \jssrev{6.60}                  & \jssrev{Rust}       & \jssrev{3.30}                  \\
\jssrev{Java}         & \jssrev{6.10}                  & \jssrev{Solidity}   & \jssrev{3.10}                  \\
\jssrev{Ruby}         & \jssrev{5.40}                  & \jssrev{C++}        & \jssrev{2.70}                  \\
\jssrev{TypeScript}   & \jssrev{3.80}                  & \jssrev{HTML}       & \jssrev{2.00}                  \\
\jssrev{Go}           & \jssrev{2.40}                  & \jssrev{CSS}        & \jssrev{1.60}                  \\ \bottomrule
\end{tabular}%
}
\end{table}

\jssrev{In the second comparison aspect, we focus on issue topics across the two platforms. Specifically, we analyzed the tags associated with bounty issues to identify the most commonly used topics on each platform. Table \ref{tab:topic} shows a comparison of the top 30 topics identified from the bounty issues from Gitcoin and Bountysource platforms. The results show that Gitcoin focuses on blockchain-related topics such as Ethereum, Blockchain, ETH, Solidity, and DeFi, which are not represented in Bountysource. On the other hand, Bountysource shows more diversity in terms of topics, with programming languages, game-related topics, and backup-related topics appearing more frequently. This difference in focus may reflect the different goals and priorities of the two platforms and suggests that issue funders and contributors may need to consider these differences when selecting a platform.}

\begin{table}[]
\centering
\caption{\jssrev{Top 30 Topics on Gitcoin and Bountysource Platforms.}}
\label{tab:topic}
\resizebox{0.4\textwidth}{!}{%
\begin{tabular}{@{}p{7.8cm}@{}}
\toprule
\jssrev{Bountysource's   top 30 topics} \\ \midrule
\jssrev{Hacktoberfest, JavaScript,   Python, Game, C, Linux, PHP, Game engine, C++, Cross-platform, RTS, Real-time   strategy, OpenRA, Command and Conquer, Strategy game engine, Backup,   Encryption, C\#, .NET, Python 3, SSH, Engine, Deduplication, Dedupe, Cython, Compression, BorgBackup, Tiberian Dawn, Red Alert, Java.} \\\midrule
                               \\\midrule
\jssrev{Gitcoin's top 30 topics}        \\\midrule
\jssrev{Ethereum, Blockchain,   Hacktoberfest, Gitcoin, Bounties, ETH, Open source, Web3, Solidity, Wallet,   Smart contracts, Dapp, Cryptocurrency, React Native, DeFi, Oracle, Android, Mobile, Bitcoin, iOS, React Native app, Messenger, ClojureScript, Clojure, Reagent, ReactNative, Re-frame, React, TypeScript, Blockchain technology.} \\ \bottomrule
\end{tabular}%
}
\end{table}

\begin{figure}[h]
        \centering
        \includegraphics[width=\columnwidth]{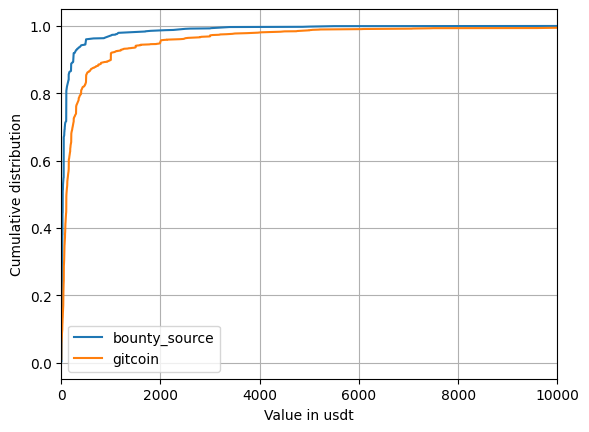}
        \caption{\jssrev{Comparison of Cumulative Distribution of USDT Value for Gitcoin and Bountysource bounty issues}}
        \label{figure:comparevalue}
\end{figure}

\jssrev{The last comparison focuses on the bounty value of issues on the two platforms. We analyze the distribution of bounty value for the successful issues of both platforms. Figure \ref{figure:comparevalue} shows the cumulative distribution of USDT value for successful issues on Gitcoin and Bountysource platforms.  By comparing the curves, it shows the differences in the distribution of USDT value for successful bounty issues on each platform. We can notice that the cumulative distribution for Gitcoin is generally lower than the cumulative distribution for Bountysource. This suggests that successful bounty issues on Gitcoin tend to have lower USDT values compared to successful bounty issues on Bountysource.}

\jssrevtwo{Moreover, since the dataset from Gitcoin (from 2017 to 2020) is newer than the Bountysource dataset from~\cite{Zhou2020} (from 2012 to 2017), there might be differences in terms of the topics, bounty values, and languages used in the bounties.
We have performed an additional study to compare the topics of the latest issues of Gitcoin and BountySource. Since the API of Bountysource is no longer available\footnote{We checked on 20th June 2023.}, we performed a manual comparison by going through the 20 latest issues on Gitcoin and Bountysource. To get the topic of the issues, we relied on multiple approaches. First, we checked the topics or labels assigned to Gitcoin or Bountysource bounties. Second, we checked the labels given to the bounties' associated GitHub issues. Third, we also checked the topics given to the GitHub project containing the issue. After analyzing the collected topics, we observed that 10 Gitcoin bounties contain blockchain-related issues compared to 3 bounties in Bountysource.} 
    
\subsection{Discussion}
\label{section-discussion}
    According to our feature analysis, correlation analysis, and feature importance analysis, we can discuss them as follows.
    
\begin{figure*}[h]
        \centering
        \includegraphics[width=0.8\textwidth]{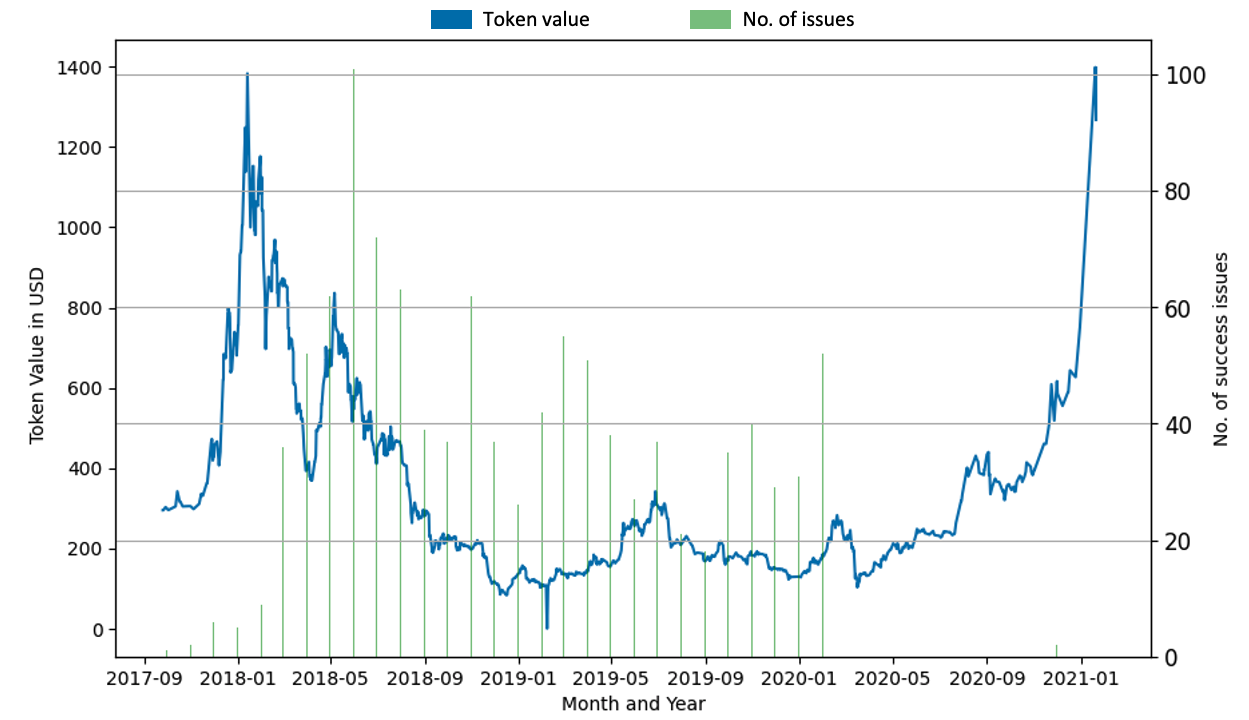}
        \caption{\jssrev{The actual value of ETH and the cumulative count of success issues}}
        \label{figure:tokenvalueETH}
\end{figure*}

    \begin{itemize}
        \item The duration in each state of an issue potentially determines the issue outcome. We found that the duration-related features, especially the duration between issues creation until the work is done, \jssrev{are highly correlated with the success of issues.} On average, those success issues usually take fourteen days to be resolved, while it is only one day for those non-success issues. In addition, the correlation analysis with the issue outcome using Point Biserial shows a strong positive correlation which indicates that the longer time spent working on issues can potentially lead to the issue being resolved and successfully have been paid. \jssrevtwo{This suggests that issue funders should carefully determine the appropriate project length when creating an issue. This should reflect the estimated duration required for issue resolution. By providing an accurate estimate, funders enable contributors to assess their ability to deliver as expected.}

        
        \item \jssrev{The length of the issue description appears to have a significant impact on the success of an issue.} Figure \ref{figure-des_issuccess} shows the number of characters of issue descriptions. As can be seen, on average, the description of the success issues is slightly longer than those of non-success. \jssrev{In addition, issues that require advanced-level experience usually have longer issue descriptions. Therefore, practitioners may benefit from spending the time and effort to write clear and detailed issue descriptions that clearly communicate the expectations and requirements for issue resolution.}
        
        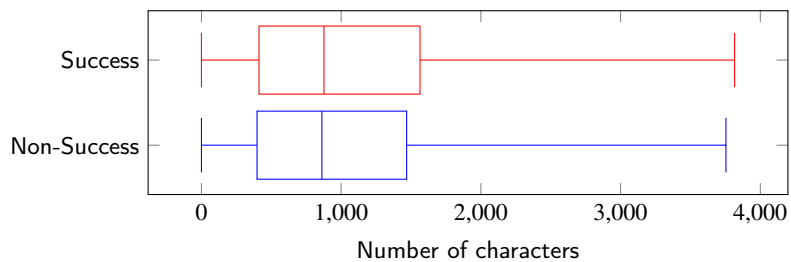
\begin{figure*}[h]
            \centering
            \input{Images/description-length-issuccess}
            \label{figure-des_issuccess}
            \caption{The distribution of description length grouped by the success and non-success issues}
        \end{figure*}
        
        

        \item \jssrev{We found that contribution type, experience level, and bounty type strongly correlate with the outcome of issues.  We then provide the bounty value of success issues in each group in Table \ref{table:avgvalueusdt}, which can help issue owners initiate the bounty value of their issues. In addition, our analysis suggests that certain token types, such as DAI and ETH, are more commonly used in successful bounty issues, indicating that these tokens may be more attractive to contributors.}

        
    \begin{table}[h!]
    \caption{The bounty value (in USD) of success issues in each feature}
    \centering
    \begin{tabular*}{\tblwidth}{@{}llrrr@{}}
    \toprule
        \multicolumn{2}{c}{\multirow{2}{*}{Group}}                    & \multicolumn{3}{c}{Value in USD}                                \\ 
        \multicolumn{2}{c}{}                                                  & Min  & Mean   & Max    \\ 
    \midrule
        \multicolumn{1}{l}{\multirow{3}{*}{\begin{tabular}[t]{@{}l@{}}Project\\types\end{tabular}}}    & Traditional & \multicolumn{1}{r}{0.00} & \multicolumn{1}{r}{148.96} & 736.48 \\ 
        \multicolumn{1}{l}{}                                   & Cooperative & \multicolumn{1}{r}{0.01} & \multicolumn{1}{r}{170.97} & 579.20 \\ 
        \multicolumn{1}{l}{}                                   & Contest     & \multicolumn{1}{r}{0.00} & \multicolumn{1}{r}{118.45} & 737.85 \\ 
        \midrule
        \multicolumn{1}{l}{\multirow{3}{*}{\begin{tabular}[t]{@{}l@{}}Experience\\levels\end{tabular}}} & Beginner    & \multicolumn{1}{r}{0.00} & \multicolumn{1}{r}{79.64}  & 500.00 \\ 
        \multicolumn{1}{l}{}                                   & Intermediate & \multicolumn{1}{r}{0.00} & \multicolumn{1}{r}{155.50} & 736.14 \\ 
        \multicolumn{1}{l}{}                                   & Advanced    & \multicolumn{1}{r}{0.10} & \multicolumn{1}{r}{200.51} & 736.48 \\ 
        \midrule
        \multicolumn{1}{l}{\multirow{3}{*}{\begin{tabular}[t]{@{}l@{}}Bounty\\types\end{tabular}}}      & Feature     & \multicolumn{1}{r}{0.00} & \multicolumn{1}{r}{146.03} & 700.00 \\ 
        \multicolumn{1}{l}{}                                   & Improvement & \multicolumn{1}{r}{0.00} & \multicolumn{1}{r}{146.14} & 691.35 \\ 
        \multicolumn{1}{l}{}                                   & Bug         & \multicolumn{1}{r}{0.00} & \multicolumn{1}{r}{94.73}  & 688.72 \\ 
    \bottomrule
    \end{tabular*}
    \label{table:avgvalueusdt}
    \end{table}
        
        
        \item \jssrev{ We found that there is a correlation between the experience level of contributors and the outcome of issues. Issues that require beginner and intermediate experience levels have a higher percentage of success compared to non-success issues. However, those issues with an advanced experience level have a higher number of non-successful outcomes. The correlation between experience level and issue outcome is negative, suggesting that the scope of work for an issue should be clear and not too complicated for the contributor to complete.}

        \item \jssrevtwo{Our study found that although the actual token value was identified as an important feature in predicting the outcomes of bounty issues in RF, it did not show a strong statistical correlation with the success issues. It is important to note that the cryptocurrency market is known for its high volatility, and the value of cryptocurrencies can fluctuate significantly. To gain further insights, we specifically focused on Ethereum (ETH), as it was the most commonly assigned token to the bounty issues in our dataset. Figure \ref{figure:tokenvalueETH} shows the actual value of ETH and the cumulative count of success issues, grouped by month and year according to the timeline. The figure shows the fluctuation in ETH value over time. Notably, during 2019, the value of ETH experienced a decline. We observed that the ETH value was particularly high in early 2018, coinciding with a high number of successful issues. However, as time progressed, both the value of ETH and the count of success issues exhibited a declining trend. These findings suggest a potential relationship between the value of cryptocurrency and the success of bounty issues.}

        \item{\jssrevtwo{Our comparative study confirms that Gitcoin's topics primarily revolve around blockchain-related issues. Consequently, the variations in programming languages can be considered a natural consequence of this focus, while the value of the bounty is comparatively lower than that of Bountysource. This could suggest issue owner to select a platform that fit their requirements. \jssrevtwo{In addition, this observation suggests that a bounty platform dedicated to a specific type of project may effectively attract attention from contributors.}}}

    \end{itemize}

\section{Threats to validity}
\label{section-tov}
    \jssrev{In this section, we discuss the threats to the validity of our study.
    \subsection{External Validity} 
    Regarding the generalizability of the work, our study focuses on only bounties issues from the Gitcoin platform. The results from our study might not be generalizable to other platforms. Therefore, our future work relates to studying issue reports from other platforms and different project settings that their bounty rewards are related to cryptocurrencies. }
    

    \jssrev{\subsection{Internal validity}} 

    \jssrev{We consider closed issues and examine only those that have been created and subsequently resolved. The data used in this analysis is sourced from the issues in Mainnet. These may have threats to the internal validity of the study. Nonetheless, we mitigate the threats by ensuring the validity of all extracted features through a validation process (e.g., manual validation process).}
    \jssrevtwo{The dataset from Gitcoin and the Bountysource dataset from~\cite{Zhou2020} overlap only in 2017. Thus, there might be a threat to validity when comparing them. We mitigate this threat by performing an additional manual comparison of the 20 latest issues in Gitcoin and Bountysource.}
    
    \jssrev{\subsection{Construct Validity}}
     \jssrev{In order to mitigate threats to construct validity, we apply the Bonferroni correction (i.e., p-value adjustment) to our correlation analysis among the extracted features and between the features and outcome of the issues. This technique helps to eliminate bias and improve the reliability of our results. We thus used statistical testing to determine the significance of any correlations found, further mitigating any potential threats to the construct validity of our experiment. In addition, to ensure the robustness of our experiments, we use a combination of techniques to analyze the correlation between the extracted features and the outcome of the issues. Specifically, we apply the Point Biserial Correlation Coefficient, which is a statistical method specifically designed for the correlation analysis of binary variables.}
     

    \jssrev{\subsection{Conclusion Validity}}
    \jssrev{We mitigate threats to conclusion validity by taking a rigorous and cautious approach when drawing conclusions based on the extracted features from the studied platform. This includes carefully limiting our conclusions to only those observations and insights that can be directly supported by the data extracted from the platform.}
    
\section{Conclusion and Future Work}
\label{section-conclusion}
    The development of open source software (OSS) projects is heavily progressed by contributions from volunteering developers. Bounty rewards in different schemes have been used to motivate participation in the OSS development. Gitcoin proposes a platform that allows issue owners to create a bounty reward using cryptocurrencies solely. The understanding of the phenomenon related to the use of bounty rewards in OSS projects promotes benefits yielded from this reward mechanism. We thus perform a study on over 4,000 Gitcoin issues by categorizing those issues into four main aspects, including primitive attributes, bounty value-related, activity-related, and duration-related features. Using statistical and machine learning techniques, we identify factors that influence the outcome of the issues. It could be served as a guide for issue owners to increase the efficiency of their bounty rewards. \jssrevtwo{We acknowledge the high volatility of the cryptocurrency market, and it is possible that fluctuations in cryptocurrency value may influence the success or failure of bounty issues. Our additional investigation into the actual value of cryptocurrency and its correlation with bounty success demonstrates the potential for a relationship. However, further study is required as part of our future work to better understand this relationship.}
    

\section{Acknowledgement}
This work (Grant No. RGNS 64-164) was financially supported by Office of the Permanent Secretary, Ministry of Higher Education, Science, Research and Innovation.













\bibliographystyle{cas-model2-names}

\bibliography{main-jss2}



\end{document}

%% file: Images/1-bounty-type-issuccess.tex
\begin{tikzpicture}
\begin{axis} [
        width = 12cm, height = 8cm, 
        xbar=0pt, xmin=0, xmax=1100,
        bar width = 10pt,
        enlarge y limits = {abs = 0.6},
        xlabel = {Number of issues},
        ylabel = {Bounty types}, 
        ytick = data,
        yticklabels = {Code Review, Design, Security, Documentation, Bug, Improvement,  Feature},
        legend pos = south east,
        reverse legend,
        nodes near coords,
        nodes near coords align = {horizontal}
    ]
    \addplot [fill=red] coordinates {
        (10, 1) (21, 2) (44, 3) (62, 4) (167, 5) (268, 6) (820, 7)
    }; 
    \addplot [fill=blue] coordinates {
        (10, 1) (25, 2) (26, 3) (162, 4) (166, 5) (411, 6) (1031, 7)
    };
    
    \legend {Non-Success Issues, Success Issues}
    
\end{axis}
\end{tikzpicture}

%% file: Images/2-project-length-issuccess.tex
\begin{tikzpicture}
\begin{axis} [
        width = 12cm, height = 6cm,
        xbar=0pt, xmin=0, xmax=1300,
        bar width = 10pt,
        enlarge y limits = {abs = 0.6},
        anchor = north,
        xlabel = {Number of issues},
        ylabel = {Project lengths},
        ytick = data,
        yticklabels = {Months, Weeks, Days, Hours},
        legend pos = south east,
        reverse legend,
        nodes near coords,
        nodes near coords align = {horizontal}
    ]
    \addplot [fill=red] coordinates {
        (17, 1) (112, 2) (506, 3) (964, 4)
    }; 
    \addplot [fill=blue] coordinates {
        (5, 1) (164, 2) (658, 3) (1285, 4)
    };
    
    \legend {Non-Success Issues, Success Issues}
    
\end{axis}
\end{tikzpicture}

%% file: Images/3-exp_level_issuccess.tex
\begin{tikzpicture}
\begin{axis} [
        width = 12cm, height = 5cm,
        xbar=0pt, xmin=0, xmax=1450,
        bar width = 10pt,
        enlarge y limits = {abs = 0.6},
        anchor = north,
        xlabel = {Number of issues},
        ylabel = {Experience levels},
        ytick = data,
        yticklabels = {Advanced, Beginner, Intermediate},
        legend pos = south east,
        reverse legend,
        nodes near coords,
        nodes near coords align = {horizontal}
    ]
    \addplot [fill=red] coordinates {
        (272, 1) (353, 2) (1010, 3)
    }; 
    \addplot [fill=blue] coordinates {
        (243, 1) (506, 2) (1405, 3)
    };
    
    \legend {Non-Success Issues, Success Issues}
    
\end{axis}
\end{tikzpicture}

%% file: Images/4-project_types_issuccess.tex
\begin{tikzpicture}
\begin{axis} [
        width = 12cm, height = 5cm,
        xbar=0pt, xmin=0, xmax=2000,
        bar width = 10pt,
        enlarge y limits = {abs = 0.6},
        anchor = north,
        xlabel = {Number of issues},
        ylabel = {\jssrev{Contribution} types},
        ytick = data,
        yticklabels = {Cooperative, Contest, Traditional},
        legend pos = south east,
        reverse legend,
        nodes near coords,
        nodes near coords align = {horizontal}
    ]
    \addplot [fill=red] coordinates {
        (245, 1) (291, 2) (1386, 3)
    }; 
    \addplot [fill=blue] coordinates {
        (170, 1) (571, 2) (1921, 3)
    };
    
    \legend {Non-Success Issues, Success Issues}
    
\end{axis}
\end{tikzpicture}

%% file: Images/9-github-comments-issuccess.tex
\begin{tikzpicture}
\begin{axis} [
        width = 12cm, height = 8cm,
        xbar=1pt, xmin=0, xmax=1300,
        bar width = 10pt,
        enlarge y limits = {abs = 0.6},
        xlabel = {Number of issues},
        ylabel = {Number of comments},
        ytick = data,
        yticklabels = {0-4, 4-8, 8-12, 12-16, 16-20, $ > 20 $},
        legend pos = north east,
        reverse legend,
        nodes near coords,
        nodes near coords align = {horizontal}
    ]
    
    \addplot [fill=red] coordinates {
        (946, 1) (318, 2) (199, 3) (142, 4) (95, 5) (222, 6) 
    };

    \addplot [fill=blue] coordinates {
        (1218, 1) (485, 2) (347, 3) (184, 4) (131,5) (297, 6) 
    };
    
    \legend {Non-Success Issues, Success Issues}
    
\end{axis}
\end{tikzpicture}

%% file: Images/10-desc-length-issuccess.tex
\begin{tikzpicture}
\begin{axis} [
        width = 12cm, height = 8cm,
        xbar=1pt, xmin=0, xmax=1500,
        bar width = 10pt,
        enlarge y limits = {abs = 0.6},
        xlabel = {Number of issues},
        ylabel = {Number of words},
        ytick = data,
        yticklabels = {0-1000, 1000-2000, 2000-3000, 3000-4000, 4000-5000, $ > 5000 $},
        legend pos = north east,
        reverse legend,
        nodes near coords,
        nodes near coords align = {horizontal}
    ]

    \addplot [fill=red] coordinates {
        (1066, 1) (453, 2) (185, 3) (100, 4) (44, 5) (74, 6) 
    };

    \addplot [fill=blue] coordinates {
        (1454, 1) (667, 2) (299, 3) (134, 4) (51,5) (57, 6) 
    };
    
    \legend {Non-Success Issues, Success Issues}
    
\end{axis}
\end{tikzpicture}

%% file: Images/5-token-names-issuccess.tex
\begin{tikzpicture}
\begin{axis} [
        width = 12cm, height = 7cm,
        xbar=0pt, xmin=0, xmax=1400,
        bar width = 10pt,
        enlarge y limits = {abs = 0.6},
        anchor = north,
        xlabel = {Number of issues},
        ylabel = {Token names},
        ytick = data,
        yticklabels = {MYB, USDC, SAI, DAI, ETH},
        legend pos = south east,
        reverse legend,
        nodes near coords,
        nodes near coords align = {horizontal}
    ]
    \addplot [fill=red] coordinates {
        (12, 1) (37, 2) (146, 3) (403, 4) (1074, 5)
    }; 
    \addplot [fill=blue] coordinates {
        (13, 1) (29, 2) (289, 3) (805, 4) (1342, 5)
    };
    
    \legend {Non-Success Issues, Success Issues}
    
\end{axis}
\end{tikzpicture}

%% file: Images/11-value_in_eth-issuccess.tex
\begin{tikzpicture}
\begin{axis} [
        width = 12cm, height = 8cm,
        xbar=1pt, xmin=0, xmax=1300,
        bar width = 10pt,
        enlarge y limits = {abs = 0.6},
        xlabel = {Number of issues},
        ylabel = {ETH},
        ytick = data,
        yticklabels = {0.0-0.2, 0.2-0.4,  0.4-0.6, 0.6-0.8, 0.8-1.0, $ > 1 $},
        legend pos = north east,
        reverse legend,
        nodes near coords,
        nodes near coords align = {horizontal}
    ]
    
    \addplot [fill=red] coordinates {
        (911, 1) (265, 2) (149, 3) (101, 4) (32, 5) (328, 6) 
    };
    
    \addplot [fill=blue] coordinates {
        (1261, 1) (444, 2) (206, 3) (117, 4) (28,5) (457, 6) 
    };
    
    \legend {Non-Success Issues, Success Issues}
    
\end{axis}
\end{tikzpicture}

%% file: Images/12-value_in_usdt-issuccess.tex
\begin{tikzpicture}
\begin{axis} [
        width = 12cm, height = 8cm,
        xbar=1pt, xmin=0, xmax=1800,
        bar width = 10pt,
        enlarge y limits = {abs = 0.6},
        xlabel = {Number of issues},
        ylabel = {USDT},
        ytick = data,
        yticklabels = {0.0-0.2, 0.2-0.4, 0.4-0.6, 0.6-0.8, 0.8-1.0, $ > 1 $},
        legend pos = north east,
        reverse legend,
        nodes near coords,
        nodes near coords align = {horizontal}
    ]

    \addplot [fill=red] coordinates {
        (1102, 1) (271, 2) (136, 3) (69, 4) (43, 5) (260, 6) 
    };
    
    \addplot [fill=blue] coordinates {
        (1717, 1) (375, 2) (187, 3) (49, 4) (31, 5) (261, 6) 
    };
    
    \legend {Non-Success Issues, Success Issues}
    
\end{axis}
\end{tikzpicture}

%% file: Images/7-all-duration.tex
\begin{tikzpicture}
    \begin{axis} [
            xmode = log,
            width = 12cm, height = 10cm,
            cycle list={blue},
            xlabel = {Number of days},
            ylabel = {Duration types},
            ytick = {1,2,3,4,5,6,7,8},
            yticklabels = {duration\_create\_to\_new\_bounty, duration\_create\_to\_start, duration\_create\_to\_stop, duration\_create\_to\_killed, duration\_create\_to\_worker\_applied, duration\_create\_to\_submitted, duration\_create\_to\_done, duration\_create\_to\_expire}, 
            boxplot/variable width,
        ]
        
        \addplot+[ 
        boxplot prepared={
                lower whisker = 0.00001, 
                lower quartile = 0.000013, 
                median = 0.0006, 
                upper quartile = 0.001, 
                upper whisker = 0.9, 
            },
        ] coordinates {};
        
        \addplot+ [
            boxplot prepared = {
                lower whisker = 0.0001, 
                lower quartile = 0.015, 
                median = 0.02, 
                upper quartile = 0.5, 
                upper whisker = 897
            },
        ] coordinates {};
        
        \addplot+ [
            boxplot prepared = {
                lower whisker = 0.0001, 
                lower quartile = 0.0003, 
                median = 0.0005, 
                upper quartile = 6.37, 
                upper whisker = 903
            },
        ] coordinates {};
        
        \addplot+ [
            boxplot prepared = {
                lower whisker = 0.0001, 
                lower quartile = 0.0003, 
                median = 0.0005, 
                upper quartile = 0.0007, 
                upper whisker = 791
            },
        ] coordinates {};
        
        \addplot+[
        boxplot prepared={
                lower whisker = 0.0001, 
                lower quartile = 0.0002, 
                median = 0.003, 
                upper quartile = 0.442, 
                upper whisker = 647.5,
            },
        ] coordinates {};
        
        \addplot+ [
            boxplot prepared = {
                lower whisker = 0.0001, 
                lower quartile = 1, 
                median = 3.78, 
                upper quartile = 8.4, 
                upper whisker = 55,
                },
        ] coordinates {};
        
        \addplot+ [
                boxplot prepared = {
                lower whisker = 0.0001, 
                lower quartile = 1, 
                median = 2.3, 
                upper quartile = 11.23,
                upper whisker = 56.98,
                },
        ] coordinates {};
        
        \addplot+ [
            boxplot prepared = {
                lower whisker = 1, 
                lower quartile = 29, 
                median = 31, 
                upper quartile = 90, 
                upper whisker = 760
            },
        ] coordinates {};
        
    \end{axis}
\end{tikzpicture}

%% file: Images/bounty-types-exp-levels.tex
\begin{tikzpicture}
\begin{axis} [
        width = 14cm, height = 11cm, 
        xbar=1pt, xmin=0, xmax=1300,
        bar width = 10pt,
        enlarge y limits = {abs = 0.6},
        xlabel = {Number of issues},
        ylabel = {Bounty types},
        ytick = data,
        yticklabels = {Code Review, Security, Design,  Feature, Documentation, Bug, Improvement},
        legend pos = south east,
        reverse legend,
        nodes near coords,
        nodes near coords align = {horizontal}
    ]
    \addplot [fill=BurntOrange] coordinates {
        (5, 1) (27, 2) (4, 3) (258, 4) (18, 5) (39, 6) (88, 7) 
    };
    
    \addplot [fill=Green] coordinates {
        (1, 1) (5, 2) (14, 3) (303, 4) (69,5) (110, 6) (158, 7) 
    };
    
    
    \addplot [fill=pink] coordinates {
        (13, 1) (37, 2) (27, 3) (1264, 4) (133,5) (177, 6) (423, 7) 
    };
    
    \legend {Advanced, Beginner, Intermediate}
    
\end{axis}
\end{tikzpicture}

%% file: Images/exp-level-worker-applied.tex
\begin{tikzpicture}
  \begin{axis}
    [
        xmode = log,
        width = 10cm, height = 5cm,
        xlabel = {Number of days},
        ylabel = {Experience levels},
        ytick={1,2,3},
        yticklabels={Beginner, Intermediate, Advanced},
    ]
    \addplot+[
    boxplot prepared={
            lower whisker = 0.0001, 
            lower quartile = 0.0002, 
            median = 0.0003, 
            upper quartile = 0.27, 
            upper whisker = 647.5
        },
    ] coordinates {}; 
    
    \addplot+[
    boxplot prepared={
            lower whisker = 0.0001, 
            lower quartile = 0.0002, 
            median = 0.002, 
            upper quartile = 0.511, 
            upper whisker = 559.4
        },
    ] coordinates {};
    
    \addplot+[
    boxplot prepared={
            lower whisker = 0.0001, 
            lower quartile = 0.001, 
            median = 0.002, 
            upper quartile = 0.22, 
            upper whisker = 630.6
        },
    ] coordinates {};

  \end{axis}
\end{tikzpicture}

%% file: Images/description-length-issuccess.tex
\begin{tikzpicture}
  \begin{axis}
    [
        width = 10cm, height = 4cm,
        xlabel = {Number of characters},
        ylabel = { },
        ytick={1,2},
        yticklabels={Non-Success, Success},
    ]
    
    \addplot+[
    boxplot prepared={
            lower whisker = 0, 
            lower quartile = 398, 
            median = 862, 
            upper quartile = 1468.75, 
            upper whisker = 3754
        },
    ] coordinates {};
    
    \addplot+[
    boxplot prepared={
            lower whisker = 0, 
            lower quartile = 412.25, 
            median = 877, 
            upper quartile = 1564.5, 
            upper whisker = 3816
        },
    ] coordinates {};

  \end{axis}
\end{tikzpicture}